\documentclass[superscriptaddress,twocolumn,10pt]{revtex4-1}

\usepackage{amsmath,graphicx,color}
\usepackage{booktabs,array,dcolumn}
\usepackage{makecell}

\usepackage{multirow,verbatim}
\usepackage{times}
\usepackage{siunitx}
\usepackage{xcolor,soul}
\newcolumntype{C}{>{\centering}p}

\usepackage{hyperref}
\usepackage{float}
\usepackage[acronym]{glossaries}
\newacronym{mlip}{MLIP}{Machine-Learned Interatomic Potential}
\newacronym{dft}{DFT}{Density Functional Theory}
\newacronym{ace}{ACE}{Atomic Cluster Expansion}
\newacronym{pes}{PES}{Potential Energy Surface}
\newacronym{aimd}{AIMD}{\textit{ab initio} Molecular Dynamics}
\newacronym{md}{MD}{Molecular Dynamics}
\newacronym{rss}{RSS}{Random Structure Search}
\newacronym{hcp}{HCP}{hexagonal close-packed}
\newacronym{bcc}{BCC}{body-centred cubic}
\newacronym{fcc}{FCC}{face-centred cubic}
\newacronym{dhcp}{dHCP}{double-hexagonal close-packed}
\newacronym{std}{STD}{Standard Deviation}
\newacronym{rmse}{RMSE}{Root Mean Square Error}
\newacronym{pbe}{PBE}{Perdew–Burke–Ernzerhof}
\newacronym{scf}{SCF}{Self-Consistent Field}
\newacronym{mp}{MP}{Monkhorst-Pack}
\newacronym{ase}{ASE}{Atomic Simulation Environment}
\newacronym{ns}{NS}{Nested Sampling}
\newacronym{eam}{EAM}{Embedded Atom Method}
\newacronym{bct}{BCT}{body-centred-tetragonal}
\newacronym{qha}{QHA}{Quasi-Harmonic Approximation}
\newacronym{xrd}{XRD}{X-Ray Diffraction}
\newacronym{fwhm}{FWHM}{Full Width at Half Maximum}
\newacronym{airss}{AIRSS}{\textit{ab initio} random structure search}
\newacronym{uspex}{USPEX}{Universal Structure Predictor: Evolutionary Xtallography}

\glsdisablehyper
\usepackage{adjustbox}
\usepackage{threeparttable}

\newif\ifnpj
\npjfalse

\begin{document}

\title{Autonomous thermodynamically informed database generation for machine-learned interatomic potentials and application to magnesium}

\author{Vincent G. Fletcher}
\email{Vincent.Fletcher@warwick.ac.uk}
\affiliation{Department of Physics, University of Warwick, Coventry, CV4 7AL, UK}

\author{Albert P. Bart\'ok}
\email{apbartok@gmail.com}
\affiliation{Department of Physics, University of Warwick, Coventry, CV4 7AL, UK}
\affiliation{Warwick Centre for Predictive Modelling, School of Engineering, University of Warwick, Coventry, CV4 7AL, UK}

\author{Livia B. P\'artay}
\email{Livia.Bartok-Partay@warwick.ac.uk}
\affiliation{Department of Chemistry, University of Warwick, Coventry, CV4 7AL, UK}

\date{\today}

\ifnpj
\maketitle
\section{Abstract}
\else
\begin{abstract}
\fi

We propose a novel approach for constructing training databases for \gls{mlip} models, specifically designed to capture phase properties across a wide range of conditions. The framework is uniquely appealing due to its ease of automation, its suitability for iterative learning, and its independence from prior knowledge of stable phases, avoiding bias towards pre-existing structural data. The approach uses \gls{ns} to explore the configuration space and generate thermodynamically relevant configurations, forming the database which undergoes \textit{ab initio} \gls{dft} evaluation. We use the \gls{ace} architecture to fit a model on the resulting database.
To demonstrate the efficiency of the framework, we apply it to magnesium, developing a model capable of accurately describing behaviour across pressure and temperature ranges of 0–600~GPa and 0–8000~K, respectively. We benchmark the model’s performance by calculating phonon spectra and elastic constants, as well as the pressure-temperature phase diagram within this region. The results showcase the power of the framework to produce robust \glspl{mlip} while maintaining transferability and generality, for reduced computational cost.
\\[4pt]
{\footnotesize UK Ministry of Defence © Crown Owned Copyright 2025/AWE}

\ifnpj
\else
\end{abstract}
\maketitle
\fi

\section{Introduction}

The development and application of machine learning-based interatomic potentials has become widespread in atomistic simulations, offering near \textit{ab initio} accuracy at a fraction of the computational cost. 
The past decade has seen rapid growth in the development of \glspl{mlip}, with the creation of different descriptors,\cite{behler_atom-centered_2011, rupp_fast_2012, bartok_representing_2013, faber_crystal_2015, thompson_spectral_2015, shapeev_moment_2016, gastegger_wacsfweighted_2018, imbalzano_automatic_2018, jain_atomic-position_2018, zhang_deep_2018, xie_crystal_2018, caro_optimizing_2019, drautz_atomic_2019, huo_unified_2022, langer_representations_2022} the use of various architectures,\cite{bartok_gaussian_2010, shapeev_moment_2016, satorras_en_2021, kocer_neural_2022, reiser_graph_2022} and the proposal of diverse workflows,\cite{janssen_pyiron_2019, allen_optimal_2022, poul_automated_2025} machine learning based models have been tailored for a wide range of materials.

While the underlying architectures of \glspl{mlip} can differ significantly, they all rely on the quality and representativeness of the training dataset. Regardless of the specific framework, the accuracy and transferability of these models are fundamentally tied to the database used for their development. 
This highlights a key avenue for advancing \glspl{mlip}: refining the construction of training datasets. In this study, we tackle the challenge by creating a procedure for constructing robust and transferable databases that capture thermodynamically relevant behaviour under a wide range of conditions, applicable to any machine learning frameworks.

Functionally, \glspl{mlip} replace computationally expensive \textit{ab initio} calculations with an approximate solution. Balancing model complexity and accuracy with computational expense, \glspl{mlip} are typically created to operate in narrow regions of phase space and are designed for each study by training on samples of the \textit{ab initio} \gls{pes}.
However, due to the high dimensionality of the \gls{pes}, and the high cost of \textit{ab initio} calculations, it is expensive to sample and unclear how to do so efficiently.
With access to vast databases and resources from years of computational studies, recent developments are pushing these frameworks to the limits by creating so-called foundation models, with a focus on sensible predictions across extensive phase and chemical space, but with reduced accuracy compared to purpose-built potentials.\cite{schutt_schnet_2018, chen_graph_2019, gasteiger_gemnet_2021, deng_chgnet_2023, choudhary_atomistic_2021, batatia_foundation_2024, neumann_orb_2024, yang_mattersim_2024, merchant_scaling_2023, kovacs_mace-off_2025, kaplan_foundational_2025}
These models can then be used as a foundation for fine-tuning by either: creating databases for a more specific application;\cite{gardner_synthetic_2024, qi_robust_2024, focassio_performance_2025} or by refitting part of the model, for greater accuracy in a specific region of phase space.\cite{novelli_fine-tuning_2024, kaur_data-efficient_2025, radova_fine-tuning_2025} Creating databases suitable to represent a high diversity of conditions has its particular challenges.
Large databases come with more \textit{ab initio} evaluations, and more data means models become more expensive to fit.
Additionally, large models are required to accurately reproduce a high diversity of properties which increases the cost of model evaluations and further increases the cost of fitting.
Another point of consideration is the importance assigned to individual sample points during the fitting procedure, to avoid artificially prioritising the accuracy of a specific phase or property, the weight associated with types of samples in the database must be taken into account during fitting.
These points highlight the importance of the density of samples within the database and, by extension, the method by which these samples are collected.

\Gls{mlip} database construction is not yet standardised, and the process of constructing a database depends both on the desired application and the available information.
If phases of the material are known, initial databases are usually constructed algorithmically, based on heuristics, such as: ground state structures, and their strained versions; surface slabs and defect configurations; and finite temperature \gls{aimd}\cite{car_unified_1985} snapshots.\cite{deringer_machine_2017,kobayashi_neural_2017,bartok_machine_2018,kruglov_phase_2019,lysogorskiy_performant_2021}
This strategy works suitably well in most cases, but it is computationally demanding, requires some prior knowledge of the material's properties, and has an inherent bias towards known and expected structures - meaning that important configurations and phases can be missed.\cite{marchant_exploring_2023}
Multiple procedures and improvements have been proposed to deal with different aspects of this problem.
For workflows that rely on \gls{aimd} to generate databases, a hybrid approach can be used, whereby energies and forces are predicted using a \gls{mlip} until some threshold of prediction uncertainty is crossed, at which point \textit{ab initio} evaluations are performed, the model retrained, and the propagation continued.\cite{stenczel_machine-learned_2023}
This type of hybrid \gls{aimd} approach, with the goal of accelerating the process of obtaining \gls{md} trajectories, can provide a 20-500 times speed-up depending on the model, phase, and material complexity.
While \gls{md} simulations are a powerful tool for suggesting thermodynamically relevant samples from within a bound region of the \gls{pes}, they are inefficient at locating a broad range of phases and rely entirely on the chosen thermodynamic conditions.
Additionally, samples are notoriously hard to decorrelate and with higher energy simulations they become more difficult to stabilise, requiring smaller time-steps to conserve energy.
To help alleviate a few of these drawbacks, some approaches run \gls{md} simulations with a \gls{mlip} but with an additional term added to the predicted energy to bias dynamics towards unseen configurations.
In doing so, \gls{md} trajectories frequently sample unseen regions of the \gls{pes}, meaning \textit{ab initio} evaluations can be limited to configurations that are not already represented in the training data.\cite{van_der_oord_hyperactive_2023, kulichenko_uncertainty-driven_2023}
This targetted approach to sampling attempts to increase sample diversity and reduce redundancy in the training database and these principles have also been applied without using \gls{md} to drive the sampling.
By leveraging local atomic environment descriptors, a measure of similarity between environments can be calculated, and sampling can then be directed to maximise descriptor diversity thereby reducing sample redundancy and promoting diversity.\cite{karabin_entropy-maximization_2020,goff_generalized_2025}
While new methods that target descriptor diversity are likely to produce representative databases, the goal of structure searching algorithms, developed for over 20 years now, has been to sample the entire structure space for stable structures, outputting a highly diverse set of relevant structures.

This need for diversity in the training data has motivated the development of methods which use structure generation algorithms to produce relevant configurations, such as \gls{uspex} or \gls{airss}.\cite{glass_uspexevolutionary_2006,pickard_ab_2011}
Such methods generate structures of a given number of atoms randomly under very general constraints -- such as a maximum volume or minimum interatomic distance or within a certain space group -- and a gradient based structure relaxation is used to find local minima of the \textit{ab initio} or \gls{mlip} \gls{pes}.
The resulting minima can then be perturbed to provide gradient information, and a \gls{mlip} trained on the resulting set of configurations.\cite{pickard_ephemeral_2022, salzbrenner_developments_2023,poul_automated_2025,kruglov_phase_2019}
Another \gls{md}-independent method includes targetted sampling along high symmetry phonon modes, which has been applied to titanium and titanium alloys.\cite{allen_optimal_2022,allen_multi-phase_2025}
These methods are highly efficient at generating large datasets with less oversight, relative to \gls{md} based procedures, but the advantage of \gls{md} simulations is that samples are drawn as a function of the Boltzmann distribution, giving a higher weight to configurations -- through increased sample density -- with the highest free energy or thermodynamic relevance.
Any sampling technique that is not based on equilibrium thermodynamics has to guide sample density to ensure all relevant phases are properly represented in the database.

In our current work, we propose the use of the \gls{ns} algorithm to combine the advantages of \gls{rss} and \gls{md} based approaches, allowing the creation of a procedure that is easy to automate, generates atomic configurations across all relevant phases, free from preconceived ideas about the materials' properties, while taking into account the thermodynamic behaviour of the material. 

The \gls{ns} algorithm generates configurations as a function of their thermodynamic relevance from the ideal gas through to the ground state structure.\cite{skilling_bayesian_2004, skilling_nested_2006, skilling_nested_2009, partay_efficient_2010, ashton_nested_2022} Each phase within the database is -- by nature of the sampling -- inherently weighted as a function of the phase-space volume it occupies. Hence, the created dataset contains structural and thermodynamic information under all thermodynamically relevant conditions representative of the entire configuration space of the material. Since only the most thermodynamically relevant configurations undergo high-cost \textit{ab initio} evaluation, our procedure decreases the cost of evaluating a database. Additionally, since the number of samples representing a basin is based on the associated phase space volume, multiple databases can be trivially combined together. This allows simple extension of the training database without the need to discard existing data or change the inherent weighting associated with specific phases or energies. 

As a test system for our procedure, we chose elemental magnesium.
Magnesium has been studied extensively across a wide pressure range ($0-100$~GPa) both experimentally,\cite{perez-albuerne_effect_1966, olijnyk_high-pressure_1985, errandonea_melting_2001, errandonea_study_2003, stinton_equation_2014} and by \textit{ab initio} calculations,\cite{moriarty_first-principles_1995, mehta_ab_2006, liu_first-principles_2009, sinko_ab_2009, li_crystal_2010} providing us with substantial benchmark data. Furthermore, extreme pressure phases have also been predicted by simulations, up to 1.6~TPa, with some recent experimental results as confirmation.\cite{gorman_experimental_2022, smirnov_comparative_2024, li_multiphase_2024}
At 0~K, the \gls{hcp} structure is the ground state crystalline phase up to approximately 53~GPa, at which point a transition to the \gls{bcc} phase is observed. \gls{bcc} remains the stable solid phase up to around 456~GPa, when a transition to the \gls{fcc} phase is predicted.
This relatively straightforward phase behaviour provides a typical scenario an interatomic potential model should be able to capture, while the vast pressure range represents a challenge for efficient database building.
The \gls{mlip} for magnesium fitted by Ibrahim \textit{et al.}\cite{ibrahim_atomic_2023} used the database generation method based on structural templates, resulting in a high quality model reliable up to 70~GPa and 4000~K trained on 40,000 individual \gls{dft} reference calculations resulting in 738,705 atomic environments.

Although the phase diagram of magnesium has been extensively investigated, some phase transitions and regions with interesting properties remain debated and this is where a high-accuracy \gls{mlip} can be critical to enhance the extent of current sampling capabilities and further our understanding of the atomic level properties of magnesium. Between 5 and 20~GPa in the high temperature solid region immediately below the melting line, the thermodynamically stable phase is debated.\cite{stinton_equation_2014} Experimental measurements suggest that an additional crystalline structure, with characteristics similar to that of the \gls{dhcp} structure, emerges but to our knowledge its precise structure has not yet been identified.\cite{stinton_equation_2014}

\section{Results and Discussion}

\subsection{Initiating and expanding a database}
\begin{figure}
\begin{center}
\includegraphics[width=8.5cm,angle=0]{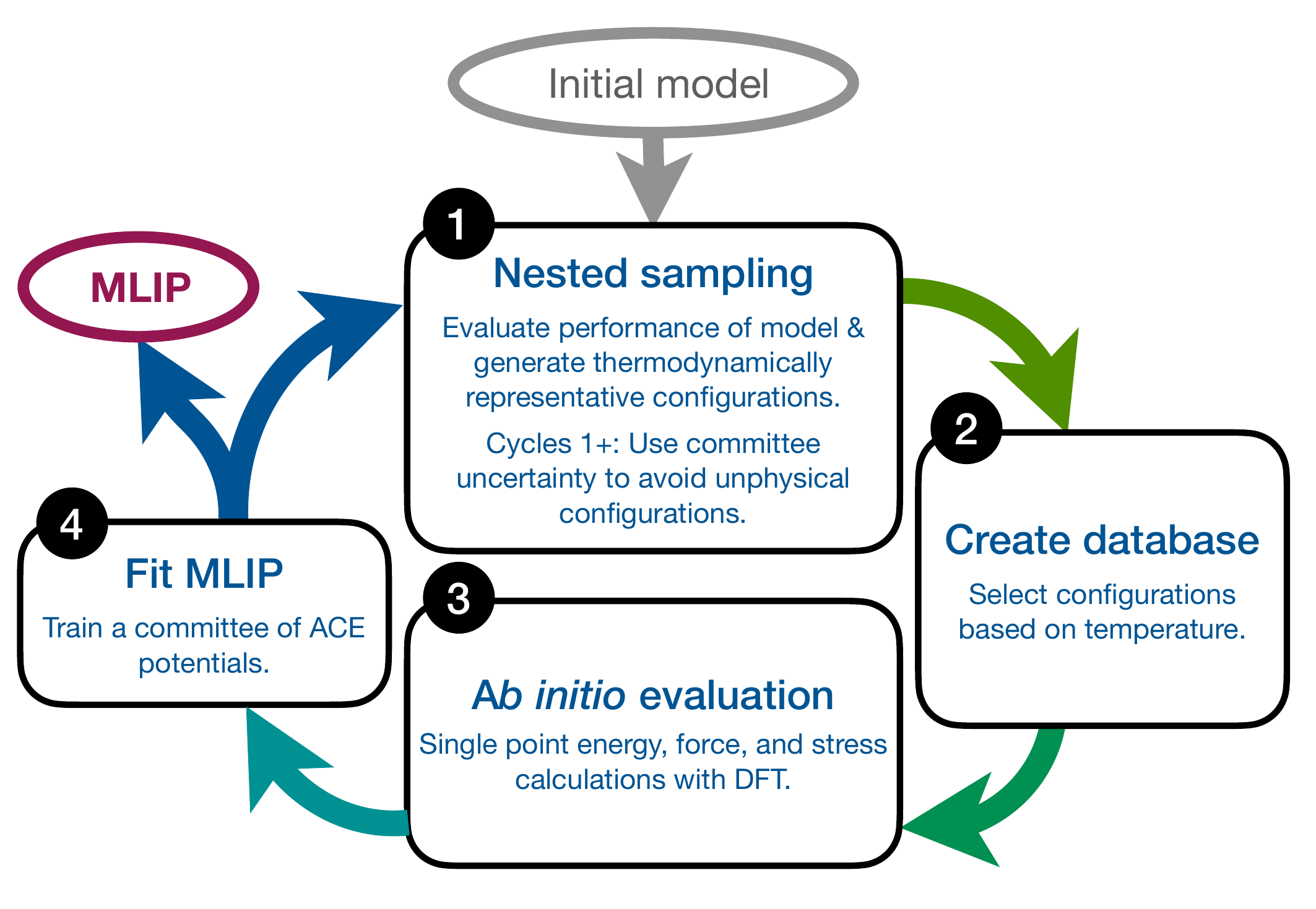}
\end{center}
\vspace{-10pt}
\caption {\textbf{Schematic workflow of the iterative potential fitting process.} The cycle starts with generating configurations from NS. In the zeroth cycle this is done using an arbitrary initial potential, an EAM in the current work. A database is autonomously constructed, guided by the thermodynamic information and samples generated by NS. The database undergoes DFT evaluation and an MLIP is (re)fitted, acting as a new input to the next cycle for further refinement if needed. We repeated this cycle 5 times.} 
\label{fig:workflow}
\end{figure}

Our method of creating the database and fitting the \gls{mlip} consists of four key stages, summarised schematically in Figure~\ref{fig:workflow}, with the key parameters for each stage given in Table~\ref{tab:key_pars}.
We refer to the produced \gls{ace} models through notation reflecting the cycle it was used in and the order and degree of the \gls{ace} model.
For example, the order 4, degree 14 \gls{ace} model used in cycle 2, will be referred to as C2~O4~D14.
First, a \gls{ns} calculation is performed using a classical interatomic potential, producing samples across all phases of the material; we then use the thermodynamic properties calculated from \gls{ns} to produce a more selective database; this refined database then undergoes \textit{ab initio} \gls{dft} evaluation; finally an \gls{ace} \gls{mlip} is fitted to the evaluated database.
We define two independent procedures for initiating or expanding a database, and since the initial cycle is performed independent of an \gls{mlip}, we label this cycle the zeroth cycle.
The specific details of these stages are discussed in the following.

\begin{table*}
\centering
\caption{\textbf{Key parameters used across the active learning cycles.} We define the critical parameters used during: the NS exploration, the selection of configurations to be added to the training database, and the subsequent MLIP fitting. The label for the models produced refer to the cycle the potential will be used in, and the order and degree of the ACE model.}
\begin{ruledtabular}
\begin{tabular}{ c|lclc|cc|cl }
      & \multicolumn{4}{c|}{Nested sampling} & \multicolumn{2}{c|}{Database Building} & \multicolumn{2}{c}{Fitting}\\
\hline
\multirow{2}{*}{Cycle} & \multicolumn{1}{c}{\multirow{2}{*}{Model used}} & \multirow{2}{*}{Atoms} & \multicolumn{1}{c}{Pressure range} & Max STD & Temp. range & \multirow{2}{*}{Samples added} & \multirow{2}{*}{Sample weight} & \multirow{2}{*}{Model produced}\\
& & & \multicolumn{1}{c}{[GPa]} & [meV/at] & [K] & & & \\
\hline
0 & EAM            & 16 & 0, 1, [5-45,5]              & N/A  & 200--3000 & 1100 & Equal        & C1 O4 D14\\
1 & ACE: C1 O4 D14 & 16 & 0, 1, [5-45,5], [60-600,20] & 62.5 & Lowest    & 39   & Equal        & C2 O4 D14\\
2 & ACE: C2 O4 D14 & 16 & 0, 1, [5-45,5], [60-600,20] & 62.5 & 200--8000 & 2801 & Equal        & C3 O4 D14\\
3 & ACE: C3 O4 D14 & 16 & 0, 1, [5-45,5], [60-600,20] & --   & 0--1000   & 390  & $\alpha=0.1$ & C4 O4 D14\\
4 & ACE: C4 O4 D14 & 8  & 0, 1, [5-45,5], [60-600,20] & --   & Lowest    & 3900 & $\alpha=0.1$ & C5 O4 D18\\
\end{tabular}
\end{ruledtabular}
\label{tab:key_pars}
\end{table*}

\subsection{Nested Sampling} \label{sec:ns_method}
The \gls{ns} method can efficiently sample high dimensional spaces and evaluate integrals of functions defined in such spaces.\cite{skilling_bayesian_2004, skilling_nested_2006, ashton_nested_2022}
\Gls{ns} has been used in a materials context to explicitly evaluate the partition functions of atomistic systems at arbitrary conditions.\cite{partay_efficient_2010}
With the full partition function, one has access to thermodynamic response functions and hence is able to determine the location of phase transitions and characterise properties of the material as a function of temperature during a post-processing step.

In general, the algorithm works by sampling the entire phase space of the material iteratively, starting from the gas phase towards the solid phase, generating configurations proportional to the phase-space that they occupy, without any prior knowledge of specific phases or structures.
The power of \gls{ns} has been demonstrated with numerous materials, from atomic clusters,\cite{partay_efficient_2010,rossi_thermodynamics_2018,dorrell_thermodynamics_2019} to soft-matter potentials,\cite{bartok_insight_2021, partay_stability_2022} and metallic systems.\cite{rosenbrock_machine-learned_2021, dorrell_pressuretemperature_2020}

Here we briefly describe the \gls{ns} technique, as employed in the current work, sampling bulk phase configurations at constant pressure, and using total-enthalpy Hamiltonian Monte Carlo to modify configurations.\cite{baldock_determining_2016}
The algorithm can be described by the following six steps:
\begin{enumerate}
    \item Generate $K$ random configurations of $N$ atoms in a cell, with a maximum volume defined by randomly generated cell vectors. These configurations are referred to as walkers.
    \item Calculate the enthalpy, $H$, of each walker and eliminate the one with the highest enthalpy.
    \item From the pool of remaining walkers, randomly select one and clone it.
    \item Perform a series of random moves on the cloned walker: cell distortions and short $NVE$ \gls{md} simulations. This process is referred to as `walking', and the number of moves is referred to as the `walk length'.
    \item Calculate the enthalpy of the walked clone. If it is lower in enthalpy than its parent, it is accepted as a new walker, if not it is rejected and step 3 onwards is repeated.
    \item Once a new walker has been accepted, the procedure from step 2 is repeated.
\end{enumerate}
The key result is that after step 2 of iteration $i$, the initial phase space volume, $\Gamma_0$, has been reduced by a known factor to $\Gamma_i$, provided the sampling is uniformly random. 
\begin{equation} \label{eqn:phase_volume}
    \Gamma_i = \Gamma_0 \exp\bigg(-\frac{i}{K}\bigg)
\end{equation}
With the change in phase-space volume at each iteration known, one can exactly compute the partition function, $Z$.
\begin{equation} \label{eqn:part_func_ns}
    Z(\beta) \approx \sum_i{\big(\Gamma_i - \Gamma_{i+1}\big)e^{-\beta H_i}}
\end{equation}
Therefore one can compute any equilibrium property of interest, $O$, as a function of the thermodynamic~$\beta$, after only one sampling procedure.
\begin{equation} \label{eqn:POI}
    O(\beta) \approx \frac{\sum_i{O_i\big(\Gamma_i - \Gamma_{i+1}\big)e^{-\beta H_i}}}{Z(\beta)}
\end{equation}

The challenge of \gls{ns} lies in producing random samples uniformly from a constantly shrinking sample space.
In practice, steps 3-5 enable this by generating a new sample configuration via a random walk - which decorrelates the clone of a randomly selected existing configuration.
These steps account for the majority of the computational cost of the algorithm; thus, in total, \gls{ns} requires on the order of $10^8$ energy evaluations for a typical system described in the current work, most of which are spent on the cloning and walking procedure.
When using \gls{ns} to calculate the pressure-temperature phase diagram of the final \gls{ace} potential, we took advantage of the recently proposed extension to the sampling, replica-exchange-\gls{ns},\cite{unglert_replica_2025} to allow better resolution of low temperature solid-solid phase transitions. 

\subsection{Initiating the Training Database}
To generate the initial magnesium database in cycle 0, a series of \gls{ns} calculations were carried out using 16-atom cells, with the interaction modelled by the \gls{eam} potential developed for magnesium by Wilson \textit{et al.}\cite{wilson_unified_2016} at eleven different pressure values: 0~GPa, 1~GPa, and every 5~GPa between 5-45~GPa inclusive.
This model underestimates the \gls{bcc} melting temperature considerably, and it also incorrectly predicts a \gls{hcp} to \gls{fcc} solid-solid transition, as shown in Figure~\ref{fig:eam_phase_diagram}. 
These shortcomings provide an ideal scenario for evaluating the ability of our training procedure to correct or expand an existing model.
While we have chosen this particular \gls{eam} model to generate initial configurations, a more approximate (e.g. Lennard-Jones) or a more advanced model (e.g. foundation \gls{mlip}) could have been selected as well.
After each \gls{ns} run, the temperature-dependent enthalpy curve was calculated using Equation~\ref{eqn:POI}, providing the temperature at which each sampled configuration has the highest probability to occur.

\begin{figure}
\begin{center}
\includegraphics[width=8.0cm,angle=0]{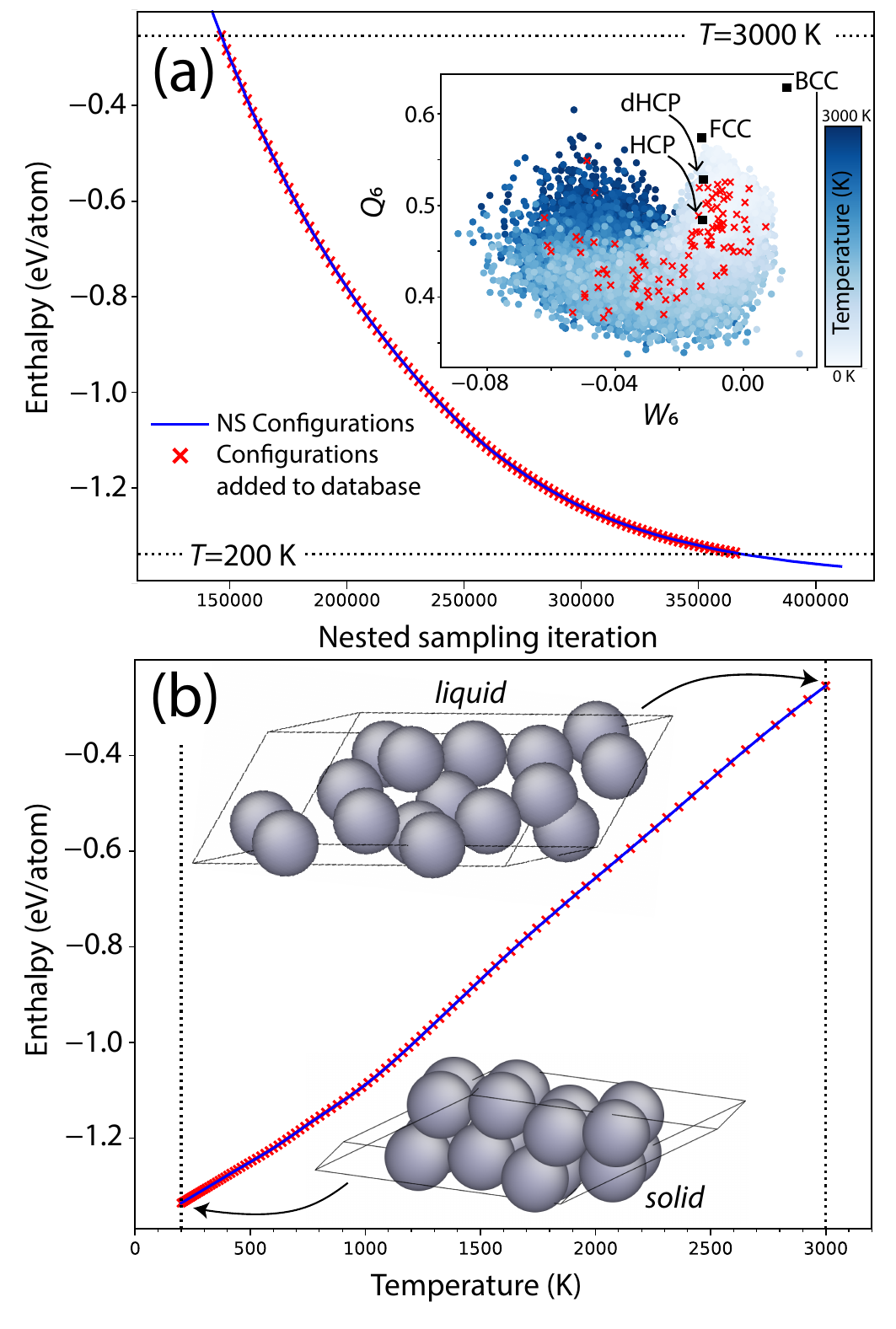}
\end{center}
\vspace{-20pt}
\caption{\textbf{The enthalpy of configurations generated during NS in cycle 0 at 1~GPa.} \textbf{a} shows the enthalpy of the samples plotted as a function of NS iterations, with the inset showing the average $Q_6$ and $W_6$ Steinhardt bond order parameters of the configurations, coloured by their associated temperature. \textbf{b} shows the enthalpy of samples plotted as a function of temperature, with snapshots of the highest and lowest enthalpy configurations selected for training shown. The 100 configurations, equally spaced in iteration number, that were added to the database are marked by red crosses. In cycle 0 all samples were generated using the EAM model.}
\label{fig:seed_db}
\end{figure}

In order to automatically exclude the least relevant gas phase configurations and select a diverse range of samples from the high-temperature liquid phase to low-temperature crystalline phases, we defined a temperature range of 200--3000~K to select configurations from.
This range generously encompasses the melting line across the entire sampled pressure range (the melting temperature of the \gls{eam} model is 1051~K and 1623~K at 1~GPa and 45~GPa, respectively).
From this range, 100 configurations were selected, equally spaced in iteration number, as shown in Figure~\ref{fig:seed_db} for the 1~GPa sampling. This provided a wide distribution of samples of and around the relevant potential energy basins as shown in Figure~\ref{fig:seed_db} through the distribution of samples across the $W_6$ and $Q_6$ Steinhardt bond order parameters.\cite{steinhardt_bond-orientational_1983}
This automatic selection was repeated at each pressure, resulting in a total number of 1100 16-atom configurations ($17{,}600$ atomic environments) collected to construct the initial database.
These configurations underwent \gls{dft} evaluation and were then used to train our first \gls{ace} potential. This model (C1~O4~D14) was then used in the consecutive training cycle: cycle 1.

\subsection{Expanding the Range}
It is trivial to repeat the previously described procedure, performing \gls{ns} calculations using the \gls{ace} model to gather more samples and expand the training set as necessary, a typical example of active learning.
However, in the early stages of this process, or when the local atomic environments deviate significantly from those represented in the training data, \glspl{mlip} can behave unpredictably.
This often manifests as so-called \emph{holes} in the \gls{pes} where the model assigns unfeasibly low energies to certain structures, typically those with unphysically short interatomic distances.
Holes are usually associated with sudden and drastic changes in the energies and forces, leading to serious issues during geometry optimisation or \gls{md} simulations. 
Such behaviour is a common challenge in \glspl{mlip}, although they can remain undetected by sampling techniques which typically explore the phase space in near-equilibrium conditions.

In contrast, due to its exhaustive sampling strategy, \gls{ns} is highly effective in uncovering these problematic regions (with increasing the number of walkers, and thus the resolution of the sampling, holes with smaller phase space volume become possible to identify).
While this capability is desirable for identifying flaws in the \gls{pes} and improving the training dataset in a targeted way, these configurations can interrupt the \gls{ns} algorithm as, once found, they dominate the rest of the sampling iterations due to their low energy.
Simply avoiding these configurations by applying a minimum distance cutoff or similar heuristics is not straightforwardly generalisable. 
For example, when describing high pressure or temperature behaviour, \emph{physical} short interatomic distances occur, which should not be removed.
Adding such configurations to the training data is unworkable due to the unfeasibility of \textit{ab initio} calculations of such configurations, which typically fail numerically due to core overlaps of the pseudopotential.

\begin{figure}
\begin{center}
\includegraphics[width=0.9\linewidth,angle=0]{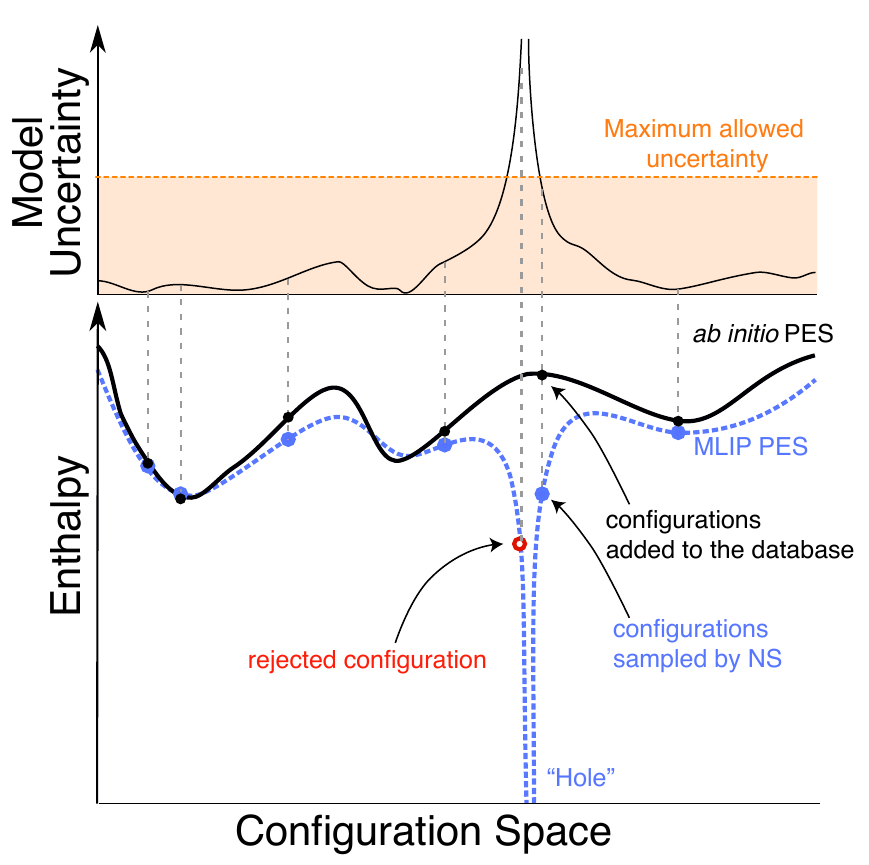}
\end{center}
\caption{\textbf{Schematic representation of sampling with a model committee STD restriction.} By sampling with the model committee STD restriction, the walkers avoid becoming trapped in holes of the PES during NS.
Solid black and dashed blue lines represent the target \textit{ab initio} PES and the MLIP PES respectively. Blue circles indicate configurations generated during the sampling, with corresponding black circles showing the same configurations after evaluation by DFT which are thereafter added to the database for MLIP training. Upper panel demonstrates the corresponding uncertainty of the model, with the orange dashed line indicating the limit, above which samples are rejected, shown by the red circle on the PES.}
\label{fig:MLIP_U_schematic}
\end{figure}

In order to employ \gls{ns} in the presence of the holes, we utilise the uncertainty quantification measure provided by the \gls{ace} committee framework, demonstrated schematically in Figure~\ref{fig:MLIP_U_schematic}.
Configurations corresponding to \gls{pes} holes contain atomic environments unseen during the fitting procedure, hence, the energy estimate of such environments have a high \gls{std}.
We found this metric to be significantly higher than for any other configuration in any other phase across the entire pressure range, and thus it is suitable to identify the \gls{pes} holes, independent of pressure and temperature.
If the \gls{std} of the committee is incorporated into the sampling as an acceptance criteria, the exploration of unfamiliar basins can be tuned to stop before the samples become unlike anything physical seen in the database which would cause the subsequent \textit{ab initio} calculations to fail.

\begin{figure*}
\begin{center}
\includegraphics[width=\linewidth,angle=0]{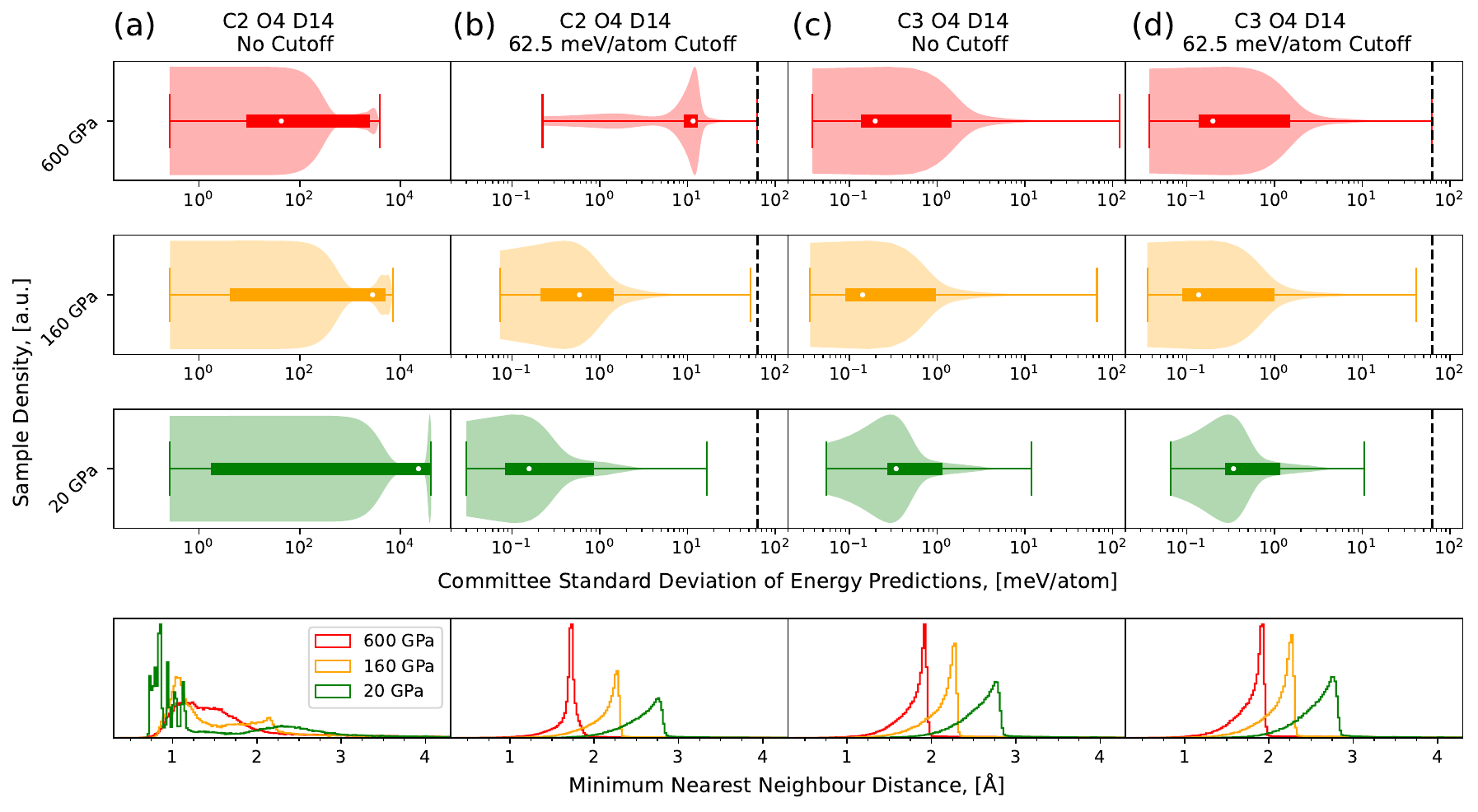}
\end{center}
\vspace{-20pt}
\caption{\textbf{Distribution of committee STD of samples and minimum interatomic distance within those samples, across cycles, across pressure, with and without the maximum STD restriction enabled.} We show how the committee STD (top three rows) and minimum bond length (bottom row) of samples changes across pressures (red: 600~GPa, orange: 160~GPa, green: 20~GPa), across cycles (cycle 2: column \textbf{a} and \textbf{b}, cycle 3: column \textbf{c} and \textbf{d}) both with (column \textbf{b} and \textbf{d}) and without (column \textbf{a} and \textbf{c}) the committee STD restriction, shown by the black dashed line. We highlight the unphysically short bond lengths and substantial STD values in column \textbf{a} and show how the STD restriction, applied in column \textbf{b}, corrects this behaviour. We also highlight the minimal effect of the STD restriction when physical samples are produced by comparing columns \textbf{c} and \textbf{d}.}
\label{fig:hole_STD}
\end{figure*}

\begin{figure*}
\begin{center}
\includegraphics[width=\linewidth,angle=0]{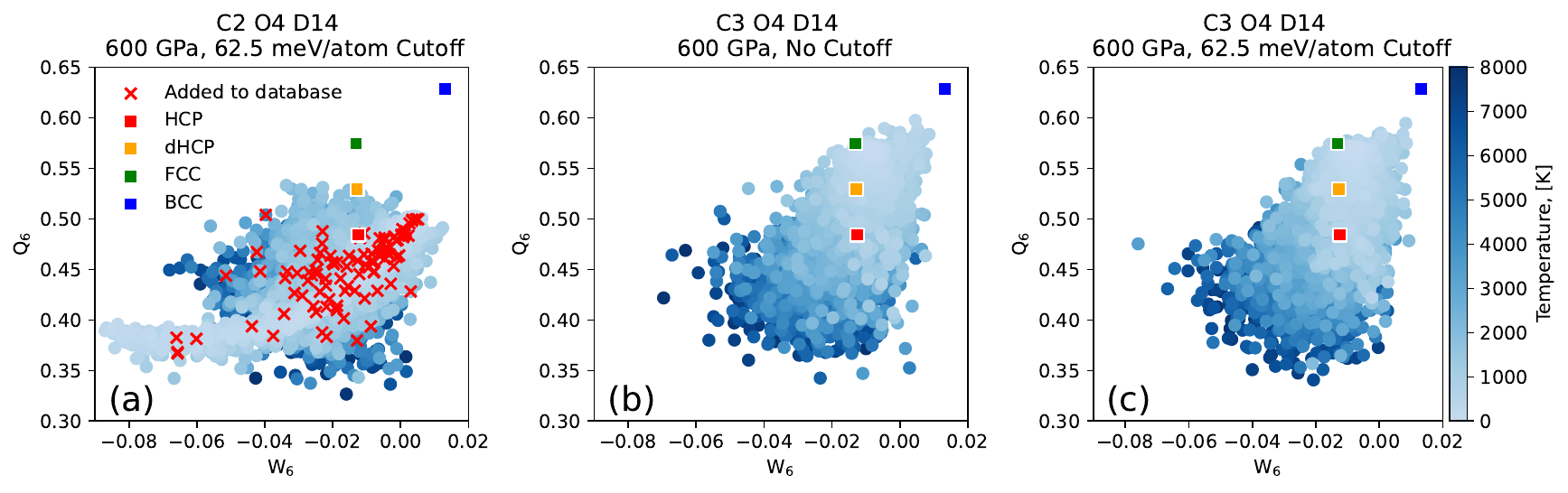}
\end{center}
\caption{\textbf{The distribution of $\boldsymbol{Q_6}$ and $\boldsymbol{W_6}$ Steinhardt bond order parameters, for the samples at 600~GPa, across cycles, with and without the maximum STD restriction enabled.} \textbf{a} shows the distribution from cycle 2 with the restriction enabled, \textbf{b} and \textbf{c} show the distributions from cycle 3 without and with the restriction enabled respectively. Samples are coloured by temperature with red crosses indicating the samples added to the database. We highlight the drastic change in distribution from cycle 2 to 3, and the minimal effect of the restriction through the similarity between \textbf{b} and \textbf{c}.}
\label{fig:uq_cut_effect}
\end{figure*}

In cycle 1, the first step in expanding the sampled pressure range is to run \gls{ns} across the entire pressure range of interest -- 0-600~GPa -- with an additional acceptance criteria introduced during the random walk to generate new samples (step 4 of the algorithm described in Section~\ref{sec:ns_method}):
after each proposed move we evaluate the committee \gls{std} in the prediction of the total energy, and if this value is above 62.5~meV/atom, the move is rejected.  
Additionally, since the presence of \gls{pes} holes makes the prediction of thermodynamic properties and temperature unreliable, which is the basis of the stopping criteria, a \gls{ns} calculation is also terminated if 90\% of the walkers have a total energy committee \gls{std} value above 60~meV/atom.
We found this criteria is only met when, during the sampling, all the walkers approach unphysically low energy regions and become immobilised.
Only the final configuration from these simulations were subject to \gls{dft} evaluation and were subsequently added to the database. Thus, only 39 16-atom configurations were added during this cycle, resulting in a total of 1139 16-atom configurations in the training database at this stage.

In the cycle 2, we repeated the procedure across the entire pressure range again. 
In case of the 1~GPa sampling, the run stopped due to high committee \gls{std}, and only the final configuration was evaluated with \gls{dft} and added to the database.
All the other \gls{ns} runs terminated when the estimated temperature reached 200~K and from the runs in the extended pressure range (60+GPa), 100 configurations were chosen that were equally spaced in iteration number, within an extended temperature range of 200-8000~K, compared to cycle 0. After this cycle, a total of 2801 16-atom configurations were added to the database.
To illustrate the use and necessity of the model uncertainty criteria, in Figure~\ref{fig:hole_STD} we show the committee \gls{std} across \gls{ns} runs at 20, 160, and 600~GPa, both with and without the \gls{std} cutoff. When attempting to perform \gls{ns} in cycle 2 without the restriction, configurations with predicted energy uncertainty up to 10~eV/atom are generated. Most of these correspond to configurations containing unphysically small interatomic distances, as also shown in Figure~\ref{fig:hole_STD}.
When the restriction of a 62.5~meV/atom cutoff is applied during sampling, unfamiliar configurations are still sampled, but they remain physical; there are no short interatomic distances or large volumes that would impede a \gls{dft} calculation. 

In cycle 3, \Gls{ns} was run across the entire pressure range of interest again. Since none of the high pressure runs stopped due to the \gls{std} stopping criteria, the \gls{std} restriction was removed.
From these \gls{ns} runs 10 configurations were selected that were equally spaced in iteration number from the final configuration up to 1000~K.
After this cycle, 390 16-atom configurations were evaluated using \gls{dft} and added to the database, for a total of 4330 16-atom configurations.
To show that the \gls{std} restriction is a suitable selective identifier of \gls{pes} holes, in column \textbf{a} of Figure~\ref{fig:hole_STD} when \gls{pes} holes are encountered the \gls{std} spikes to around 10~eV/atom but in column \textbf{c} when the holes have been fixed the \gls{std} only spikes to around 0.1~eV/atom. Additionally, to demonstrate that the restriction does not affect the sampling when \gls{mlip} \gls{pes} holes are not encountered, Figure~\ref{fig:uq_cut_effect} shows the distribution of samples in $W_6$ - $Q_6$ parameter space when a hole is encountered and no restriction is applied (a), when a hole is not encountered and the restriction is not applied (b), and when a hole is not encountered and the restriction is applied (c). The indistinguishable differences between distributions \textbf{b} and \textbf{c} support the use of this restriction during sampling.

While subsequent \gls{ns} simulations did not find more unphysical configurations, additional configurations did improve our benchmarks metrics.
In this stage of the active learning process we weighted our samples according to Equation~\ref{eqn:reweight}, thus lower enthalpy configurations have greater importance. To avoid overweighting the low pressure configurations, the weights were rescaled at each sampled pressure.
This weighting scheme also ensures that high-energy and unphysical configurations have lower weights associated with them, as accuracy in the corresponding regions of the \gls{pes} is less important.
We would like to emphasise, that our procedure does not guarantee that all holes are eliminated; while unbiased and exhaustive \gls{ns} exploration has not identified further unphysical regions, there is a possibility that a higher resolution sampling (i.e. employing more walkers), a larger system size, different thermodynamic conditions, or with a more flexible \gls{mlip} model, further \gls{pes} holes could be sampled.

Cycle 3 has produced an excellent general purpose \gls{mlip}, C4~O4~D14, capturing the expected thermodynamic properties of magnesium as shown in detail in the results section.
For cycle 4, in order to improve the prediction of low-enthalpy microscopic properties, we performed a fourth cycle of our procedure. For computational efficiency, and to be able to evaluate more samples concentrating on low entropy phases, \gls{ns} was performed with 8 atoms.
The final 100 configurations from each sampled pressure were evaluated with \gls{dft} and added to the database. 
Our final database contains 8230 configurations with $100{,}480$~atoms in total.

\subsection{0~K Enthalpy Curves and Isotropic Volume Expansion}
\begin{figure*}
\centering
\includegraphics[width=0.95\textwidth,height=0.95\textheight]{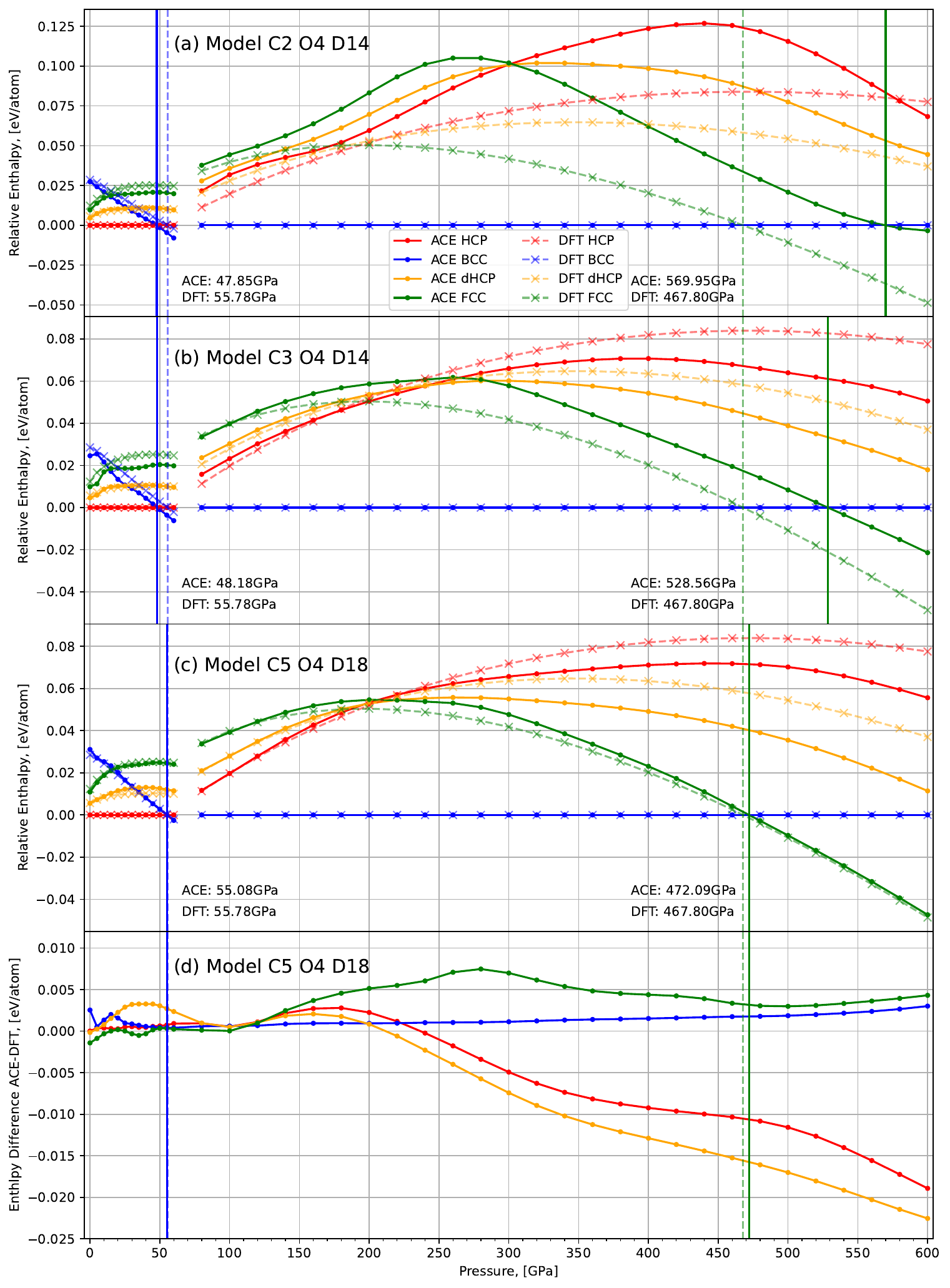}
\caption{\textbf{The 0~K enthalpy minima of the four key crystal structures of magnesium, up to 600~GPa, for three different ACE models compared to DFT.} Panels \textbf{a}, \textbf{b}, and \textbf{c} use the ACE models used to generate samples in cycle 2, 3 and 5 respectively. \textbf{d} shows the enthalpy differences between the ACE and DFT predictions in \textbf{c}. Vertical lines indicate a change of the ground state structure. Note that at 60~GPa, due to the change in the ground-state structure, we change from the HCP reference to the BCC reference creating a discontinuity in how the lines are shown.}
\label{fig:enth_cur}
\end{figure*}

To benchmark the ability of the \gls{mlip} to predict the relative stability of different crystal structures, and thus to identify the ground state, we calculated the enthalpy at 0~K for \gls{bcc}, \gls{fcc}, \gls{hcp}, and \gls{dhcp} structures. Figure~\ref{fig:enth_cur} shows these results in the 0-600~GPa pressure range, obtained at different stages of our active learning procedure.
We would like to emphasise that none of these crystalline structures have been manually added to the database during the training and while \gls{ns} performed with the \gls{eam} potential in cycle 0 sampled a range of relevant solid structures, these provided limited and often incorrect lattice parameters.
Despite this, the C2~O4~D14 potential trained at the end of cycle 1 already provides reasonable predictions of relative phase stabilities (Figure~\ref{fig:enth_cur} panel (a)), particularly at low pressures, and while the relative enthalpy difference between \gls{dft} and the \gls{ace} potential deviates more at extreme pressures, the relative stability order between phases is already correct.
This is a testament to both the ability of the \gls{ace} architecture to accurately interpolate, based on only a small amount of data, and the quality of data collected through our procedure.

As the high pressure phases are sampled more extensively during cycles 3 and 4, enthalpy predictions improve. However, we found that significant improvements can only be achieved by increasing the flexibility of the model, rather than by additional samples, and thus we increased the degree of our model (C5~O4~D18) - increasing flexibility and computational cost - and refitted to the final database.
This new model provides excellent agreement with the \gls{hcp} to \gls{bcc}, and \gls{bcc} to \gls{fcc} phase transition pressures. 

One might also notice that the enthalpy predictions of the high-pressure metastable phases, \gls{hcp} and \gls{dhcp}, are less accurate than those involved in the phase transition, \gls{fcc} and \gls{bcc} (Figure~\ref{fig:enth_cur} panel (d)).
This naturally emerges from our database building procedure, as \gls{ns} samples the most thermodynamically relevant basins, hence the metastable structures are less well represented in the training data.
Selectivity towards thermodynamically relevant basins increases computational efficiency when constructing databases and avoids the evaluation of unimportant configurations.

\begin{figure}
\begin{center}
\includegraphics[width=8.5cm,angle=0]{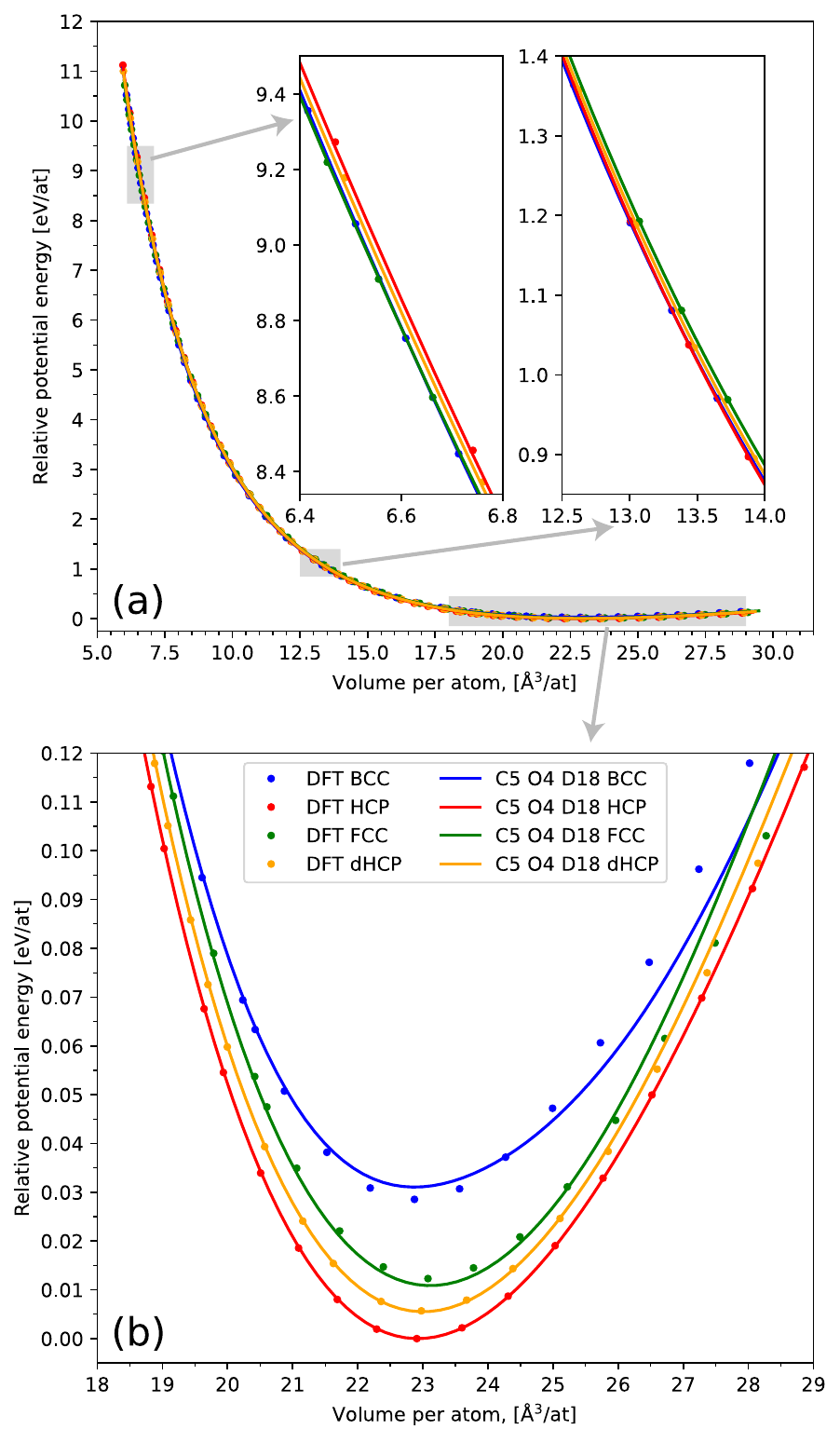}
\end{center}
\vspace{-20pt}
\caption{\textbf{Energy as a function of volume for the four key crystal structures of magnesium, calculated using DFT and our final ACE model, C5~O4~D18.} DFT results are shown as single points, while model predictions are shown as solid lines. \textbf{b} shows the minimum energy region of the curves. Note, the cells were expanded isotropically so for HCP and dHCP, other than at 0~GPa, these are not related to the enthalpy minima.}
\label{fig:volume_energy}
\end{figure}

We also present the potential-energy minima isotropic volume expansion curves, shown in Figure~\ref{fig:volume_energy}, as a demonstration of the functionally smooth local minima, absent of any potential-energy holes up to very small cell volumes. We also compare these results to those obtained via \gls{dft} of the same structures and observe excellent agreement for the lowest energy phases on the left-hand side of the graph. As explained above, we do not expect perfect results from phases that are not thermodynamically relevant, such as a high-volume \gls{bcc} crystal, and since the \gls{ns} calculations generating the training data were not performed at negative pressure values, it is expected that on the right-hand side of the curve, corresponding to high-volume crystal structures, the predictions will be poor. The results remain physical despite the lack of data.

\subsection{BCC-FCC Transition Pathway (Bain Path)}
In order to benchmark the behaviour of the potential in regions of phase space that are important for determining the mechanics of phase transitions, we evaluated the transition pathway between \gls{bcc}-\gls{fcc} phases, known as the Bain path, shown schematically in Figure~\ref{fig:bain_path}.

\begin{figure}
\begin{center}
\includegraphics[width=8.5cm,angle=0]{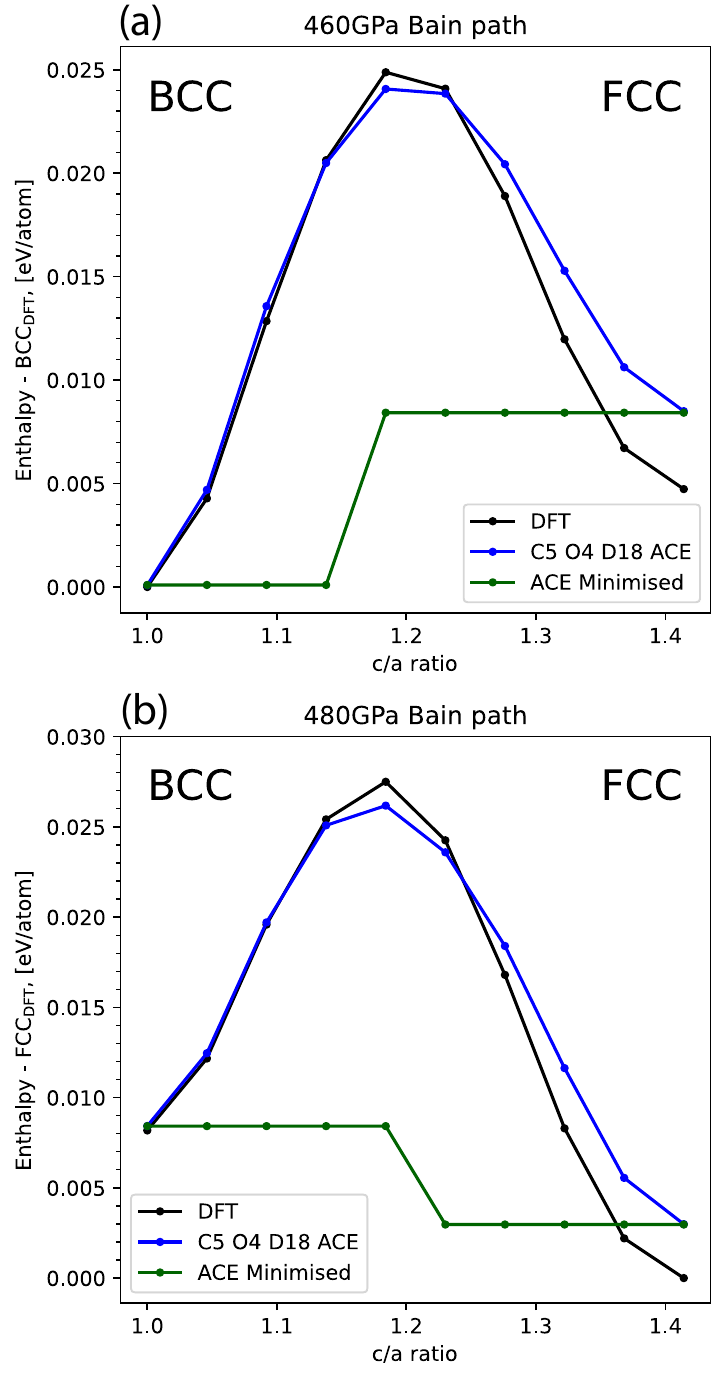}
\end{center}
\vspace{-20pt}
\caption{\textbf{The enthalpies of BCT intermediate structures along the Bain path at different pressures.} \textbf{a} and \textbf{b} show the results at 460~GPa and 480~GPa respectively. DFT results are shown in black with C5~O4~D18 ACE model predictions in blue. We minimised these intermediate structures using the ACE model, to show no spurious minima along the pathway exist, and plot the enthalpies of these minimised structures in green. The Bain path is shown schematically in Figure~\ref{fig:bain_path}.} 
\label{fig:bain_path_res}
\end{figure}

The results in Figure~\ref{fig:bain_path_res} show the Bain path at two different pressures, one where the \gls{bcc} phase is the most stable and at higher pressure when \gls{fcc} is the ground state. These show that the enthalpy of the \gls{fcc} phase is overestimated, which was also observed from the ground state enthalpy comparison plots (Figure~\ref{fig:enth_cur}), but the \gls{bcc} phase is in excellent agreement with the \gls{dft} results at both pressures. Due to the slight underestimation of the transition point between the phases and the overestimation of the \gls{fcc} trough, it seems to suggest the model better fits to the surface at finite temperatures rather than 0~K.

\subsection{Phonons}
The phonon spectra are representative of how well the \gls{mlip} can reproduce the forces within potential energy basins in specific high-symmetry directions. 
This benchmark can be challenging as it depends on the gradients of the potential energy landscape, which requires dense sampling around the minimum to correctly approximate. Given we do not explicitly supply the basin minima, or direct sampling along high-symmetry directions, this benchmark could be particularly challenging.

\begin{figure*}
\begin{center}
\includegraphics[width=\linewidth]{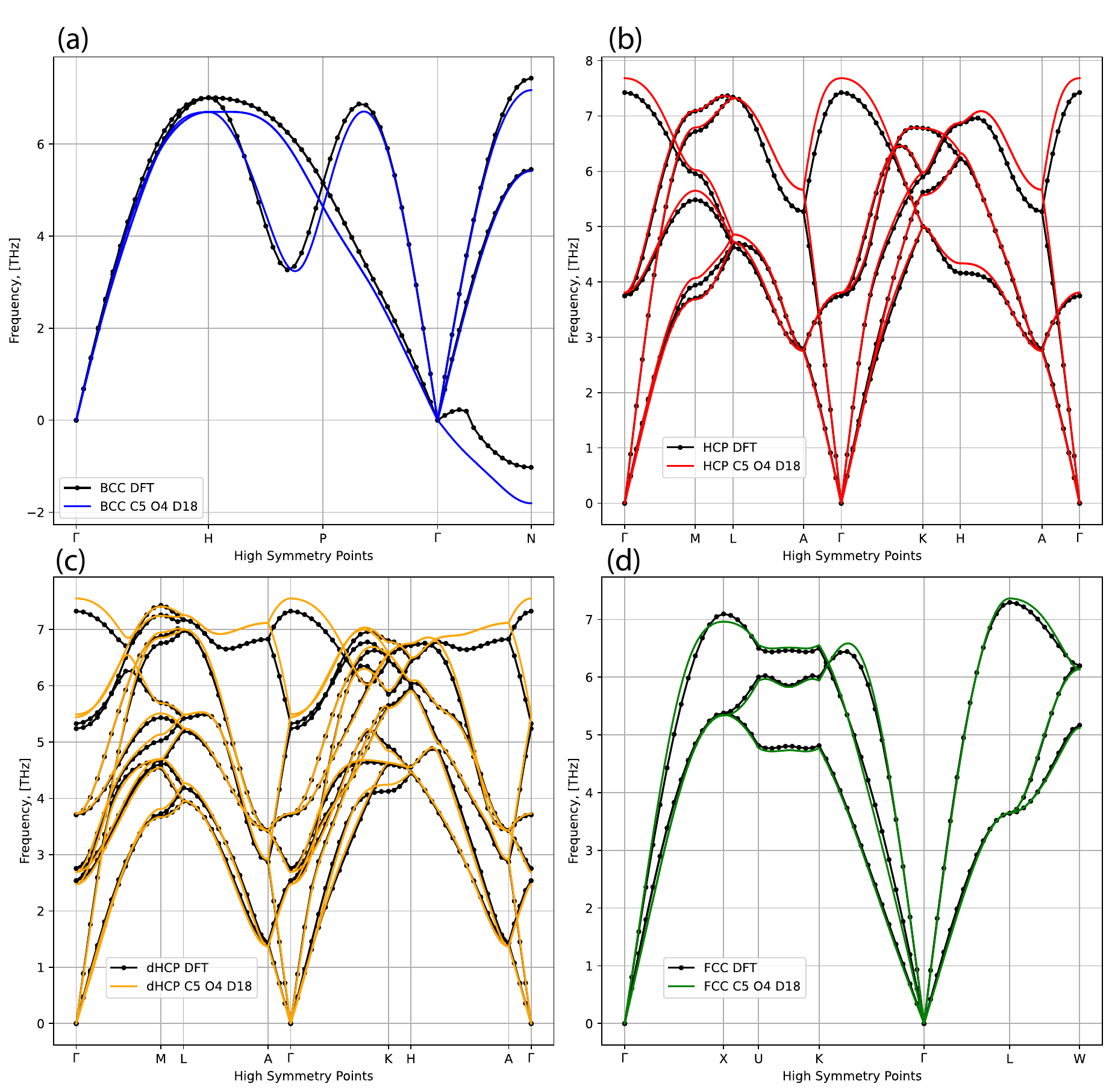}
\end{center}
\vspace{-20pt}
\caption{\textbf{Phonon spectra for the four key crystal structures at 0~GPa.} Spectra were calculated using DFT and our final ACE model, C5~O4~D18. DFT results are shown in black, with ACE results given for BCC in blue (\textbf{a}), for HCP in red (\textbf{b}), dHCP in orange (\textbf{c}), and for FCC in green (\textbf{d}).}
\label{fig:phonons}
\end{figure*}

We present the phonon spectra at 0~GPa in Figure~\ref{fig:phonons}, where we demonstrate good agreement with the \gls{dft} benchmarks across all four crystal structures identified in the current work. We also detect softening of the unstable \gls{bcc} phonon mode, even though \gls{bcc} is not the lowest enthalpy phase at 0~GPa and subsequently not well represented in the database.

\begin{figure*}
\begin{center} 
\includegraphics[width=\linewidth,height=\textheight]{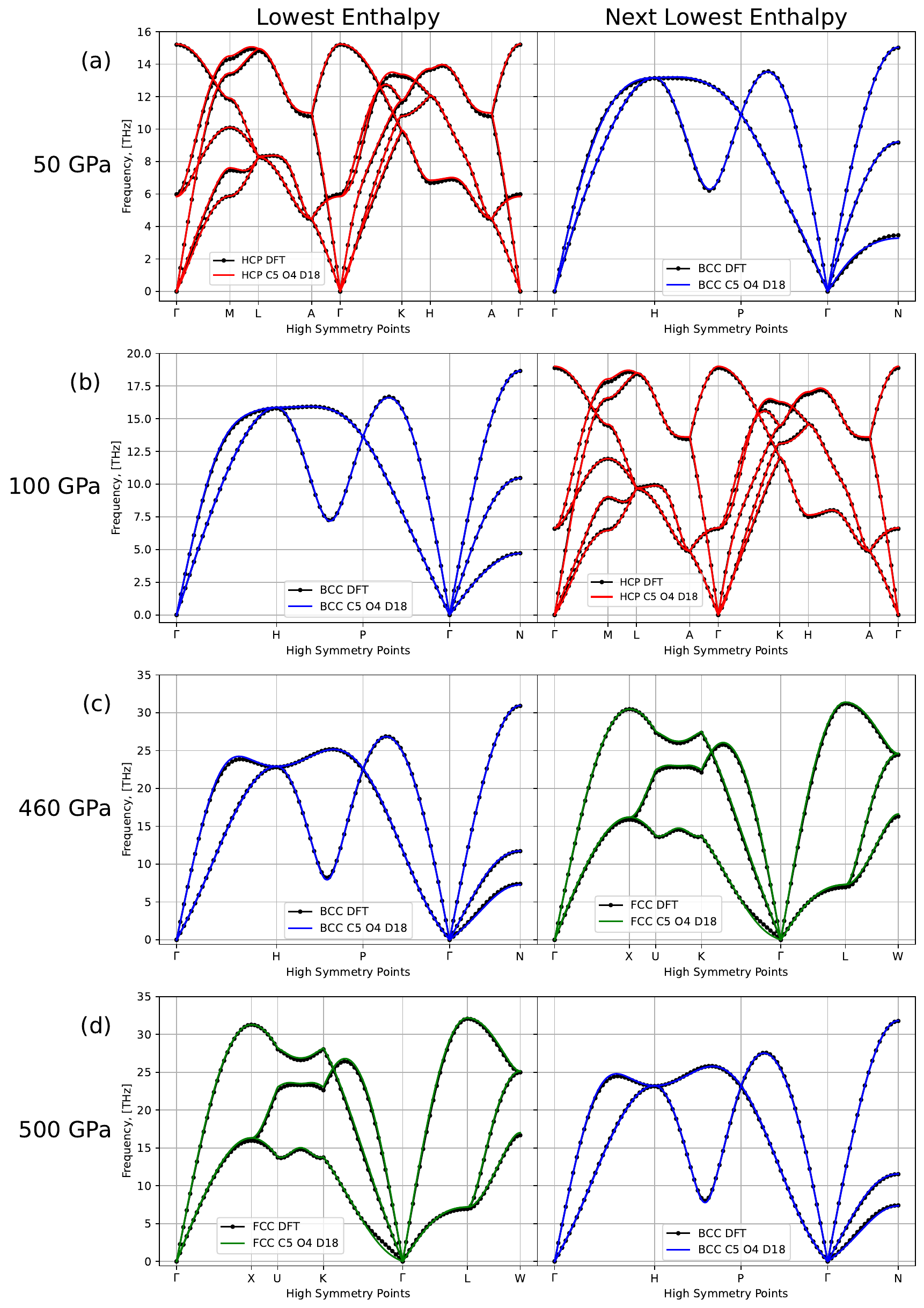}
\end{center}
\vspace{-20pt}
\caption{\textbf{Phonon spectra for FCC, BCC, and HCP, around the 0~K phase transition pressures.} Spectra were calculated using DFT and our final ACE model, C5~O4~D18. With the HCP-BCC and BCC-FCC ground state transitions occurring at 56~GPa and 468~GPa respectively, we compare the HCP (red) and BCC (blue) phonons at 50~GPa (\textbf{a}) and 100~GPa (\textbf{b}), and the BCC (blue) and FCC (green) phonons at 460~GPa (\textbf{c}) and 500~GPa (\textbf{d}), to DFT (black).}
\label{fig:hp_phonons}
\end{figure*}

Compared to previous studies,\cite{ibrahim_atomic_2023} we are not expecting a uniformly excellent agreement across the phases, as thermodynamically unstable phases are undersampled or not sampled at all in our approach.
We demonstrate, however, that phonon dispersions at finite pressures, as presented in Figure~\ref{fig:hp_phonons}, show excellent agreement with the \gls{dft} reference for the thermodynamically stable phase and the next lowest enthalpy phase across a broad range of pressure values.
This result indicates our procedure is working as expected from the sampling properties of \gls{ns} and that our \gls{mlip} can be trusted to reproduce difficult properties of the  material under near-equilibrium conditions.

\subsection{Elastic Constants}
The elastic constants provide a measure of the accuracy of the stresses predicted by our \gls{mlip} at 0~GPa. We present the elastic constants for the four principal crystal structures of magnesium compared to \gls{dft} in Table~\ref{tab:elastics}.
At 0~GPa the most stable phase is \gls{hcp} and therefore the most sampled by \gls{ns} and thus highly accurate, but the elastic constants for the other metastable solid phases also show excellent agreement, with the largest differences only on the order of 5~GPa compared to \gls{dft}.

\begin{table}
\centering
\caption{\textbf{Key components of the elastic constant matrix of the four key structures, calculated using DFT and our final ACE model, C5~O4~D18.} All units are in GPa.}
\begin{ruledtabular}
\begin{tabular}{ c|cccccccc }
\multirow{2}{*}{Component} & \multicolumn{2}{c}{HCP} & \multicolumn{2}{c}{dHCP} & \multicolumn{2}{c}{FCC} & \multicolumn{2}{c}{BCC}\\
    & DFT & ACE & DFT & ACE & DFT & ACE & DFT & ACE\\
\hline
C11 & 64.61 & 63.82 & 63.22 & 60.25 & 43.47 & 47.05 & 33.54 & 24.67 \\
C12 & 22.97 & 23.07 & 23.59 & 24.78 & 31.11 & 30.36 & 35.99 & 29.61 \\
C13 & 21.16 & 20.79 & 20.69 & 21.22 &  --   &  --   &  --   &  --   \\
C33 & 64.68 & 68.62 & 64.59 & 68.11 &  --   &  --   &  --   &  --   \\
C44 & 17.59 & 19.04 & 15.39 & 14.88 & 23.76 & 18.49 & 29.85 & 31.34 \\
\end{tabular}
\end{ruledtabular}
\label{tab:elastics}
\end{table}

\subsection{Phase Diagram}
\begin{figure}
\begin{center}
\includegraphics[width=8.5cm,angle=0]{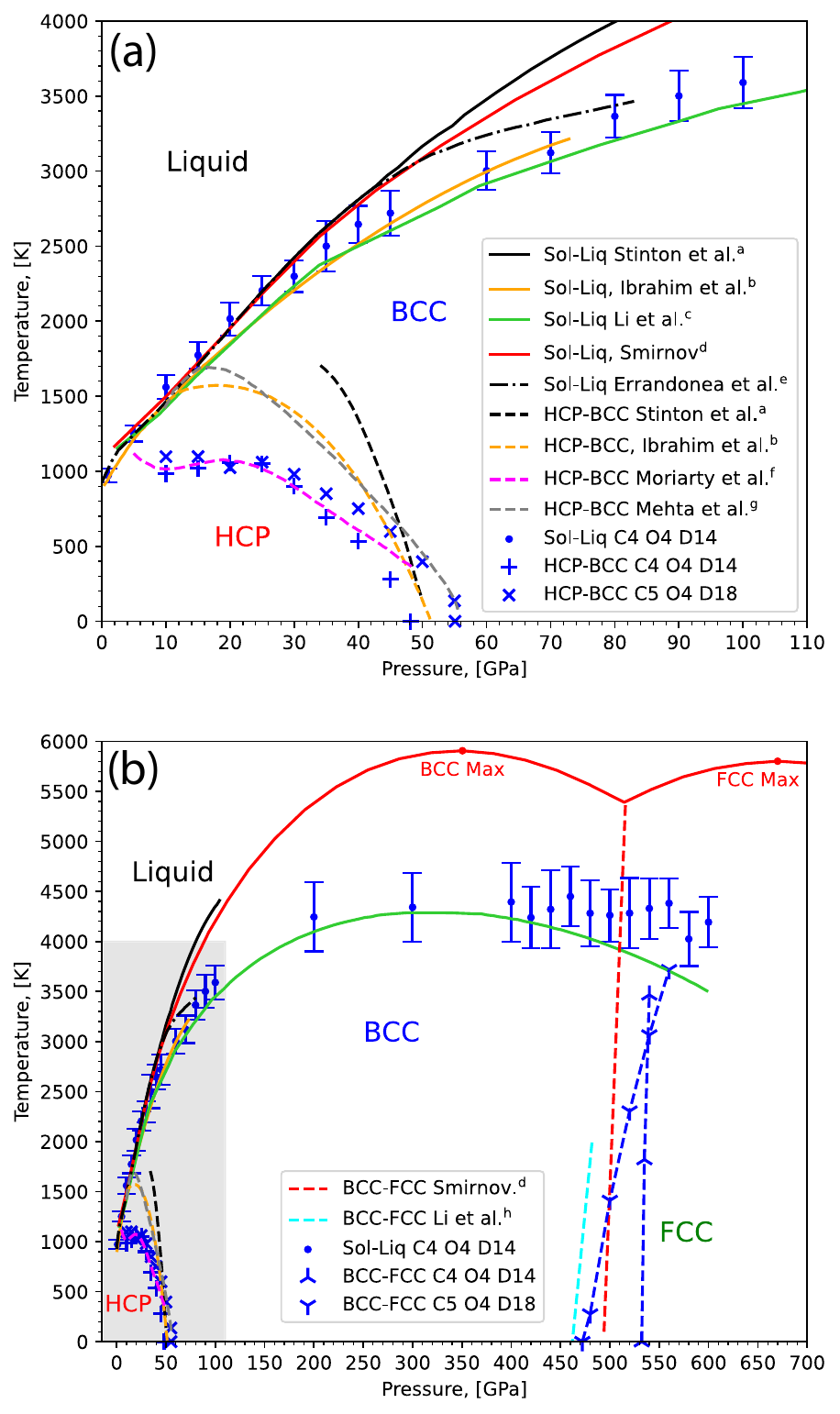}
\end{center}
\vspace{-20pt}
\caption{\textbf{Pressure-temperature phase diagram of magnesium.} We compare predictions evaluated using the C4~O4~D14 ACE potential developed in the current work (blue symbols) to previous experimental measurements and computational predictions (a:\cite{stinton_equation_2014}, b:\cite{ibrahim_atomic_2023}, c:\cite{li_multiphase_2024}, d:\cite{smirnov_comparative_2024}, e:\cite{errandonea_melting_2001}, f:\cite{moriarty_first-principles_1995}, g:\cite{mehta_ab_2006}, h:\cite{li_crystal_2010}). The error bars on our NS results represent the full-width half-maximum of the calculated constant pressure heat capacity peaks (discussion on this can be found in Appendix~\ref{sec:error_bar}). Panel \textbf{a} shows the lower pressure grey region of the phase diagram in panel \textbf{b}, enlarged.}
\label{fig:phase_diagram}
\end{figure}

To comprehensively evaluate the phase behaviour of our \gls{ace} model, we calculated the phase diagram across a wide pressure range of $1–600$~GPa as shown in Figure~\ref{fig:phase_diagram}. We employed \gls{ns} to perform an unbiased exploration of the configurational space, ensuring that all relevant phases are considered and that no erroneously stabilised structures influence the results.
Through \gls{ns} our \gls{mlip} predicts the experimentally expected phases of magnesium across a very wide pressure range, without failures due to holes in the \gls{pes}.
All results are in close agreement with existing data where available, further validating the predictions made by our model.

First, looking at the melting line between 1-45~GPa, there is excellent agreement with both existing experimental data and other computational studies. Beyond 50~GPa, our results, and those of other computational studies, agree best with those reported by Errandonea \textit{et al.},\cite{errandonea_melting_2001} in contrast to those reported by Stinton \textit{et al.}.\cite{stinton_equation_2014} 

At high pressures, where experimental melting data has not been collected, our results agree very closely with recently published \textit{ab initio} results from Li \textit{et al.}\cite{li_multiphase_2024} for the \gls{bcc} melting line up to 400~GPa.
We note that our predicted melting temperatures are consistently above those of Li \textit{et al.} but within 200~K.
\Gls{ns} results typically suffer from finite size effects that push transition temperature predictions above their ideal limit, so this trend is to be expected, and our results could be further improved by incorporating finite size effect corrections.
The error bars correspond to the width of the heat capacity peaks, whose broadening is a direct result of the finite size effect due to the system size of 64 atoms. This is further discussed in Appendix~\ref{sec:error_bar}.

Past 400~GPa as pressures approach the \gls{bcc}-\gls{fcc} solid-solid transition, our results begin to deviate from those of Li \textit{et al.}\cite{li_multiphase_2024} -- which is expected since Li \textit{et al.} did not consider the \gls{fcc} phase in this region.
The melting line above 400~GPa becomes relatively flat, similar to what has been predicted computationally both by Li \textit{et al.} and Smirnov,\cite{smirnov_comparative_2024} although the transition temperatures in the latter are considerably higher. 

\begin{figure}
\begin{center}
\includegraphics[width=8.5cm,angle=0]{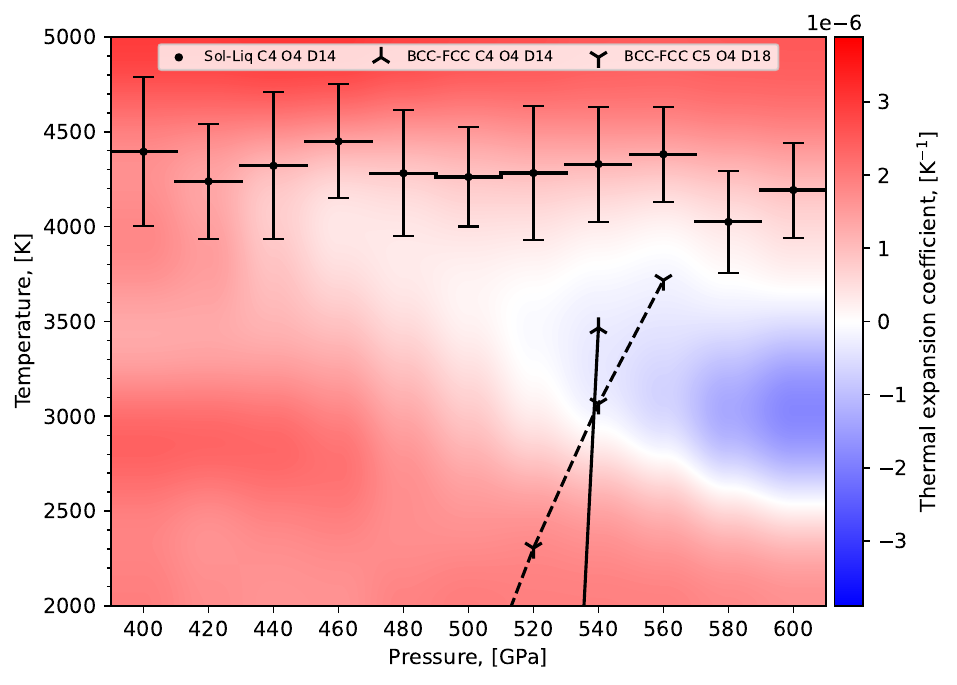}
\end{center}
\vspace{-20pt}
\caption{\textbf{Thermal expansion coefficients between 400 and 600~GPa.} Calculated from the \gls{ns} runs with our C4~O4~D14 ACE model. The melting points are displayed as horizontal black bars and the QHA prediction for the BCC-FCC solid-solid transition is shown  with triangular black markers. The red regions show the typical behaviour of volume increasing with increasing temperature, while the expanding and darkening blue region with increasing pressure demonstrate negative thermal expansion. Note to remove some of the noise resulting from differentiation of stochastic averages, we have applied Gaussian filtering with widths of 200~K and 20~GPa on the temperature and pressure axes, respectively. The unfiltered data is provided in Appendix~\ref{sec:lh_and_enths}.}
\label{fig:thermal_exp}
\end{figure}

To gain further insight into the character of the melting line, we determined the thermal expansion coefficients across the \gls{ns} runs, which is shown in Figure~\ref{fig:thermal_exp}.
Across the \gls{ns} runs from 400-440~GPa there is positive thermal expansion up to 460~GPa where the thermal expansion becomes negative indicating a maximum of the \gls{bcc} melting line within the pressure range of 440-460~GPa.
Thermal expansion remains negative, and becomes more so, up to the final measured pressure of 600~GPa.

\begin{figure}
\begin{center}
\includegraphics[width=8.5cm,angle=0]{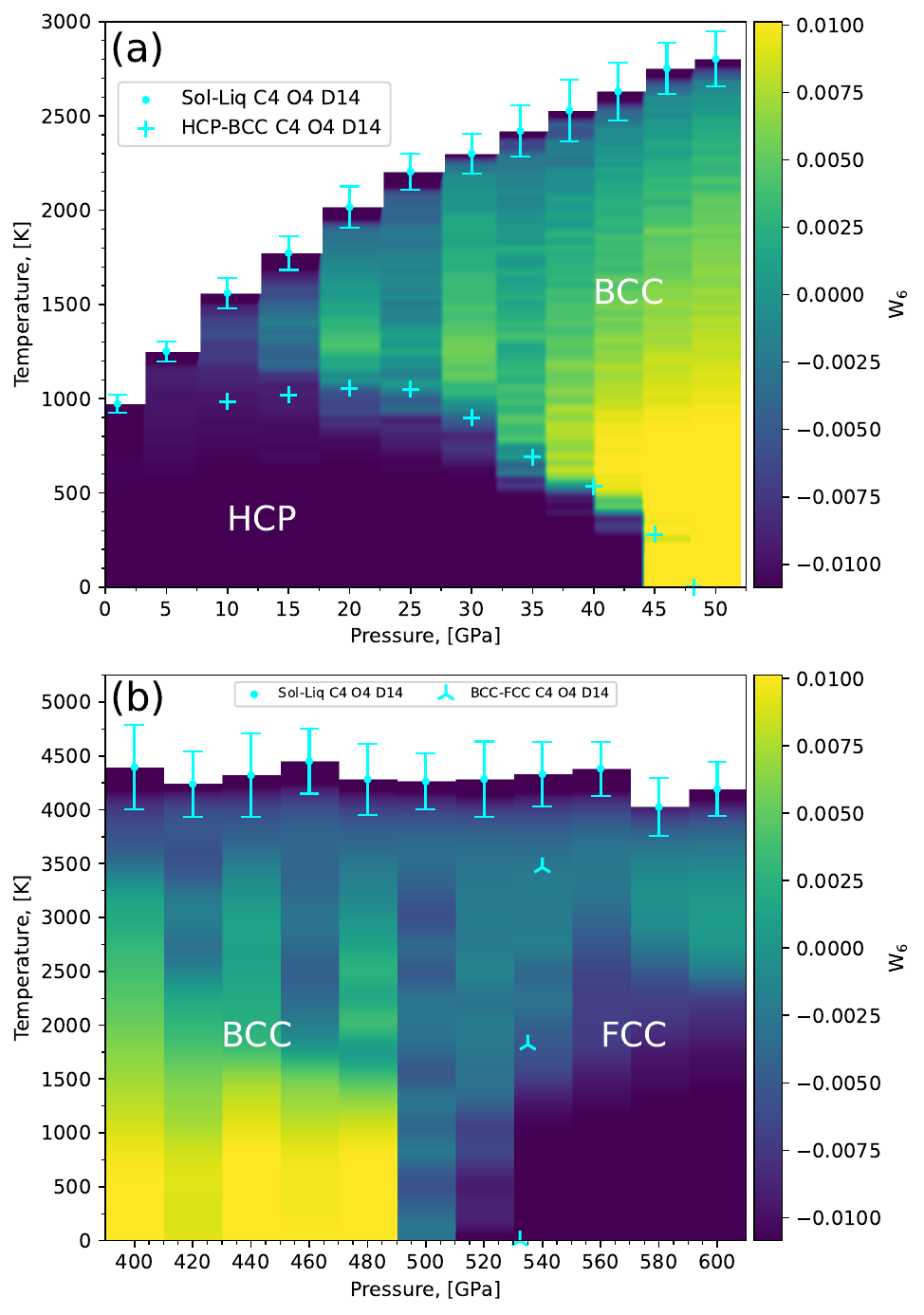}
\end{center}
\vspace{-20pt}
\caption{\textbf{$\boldsymbol{W_6}$ Steinhardt bond order parameters around predicted solid-solid transitions on the phase diagram.} Configurations were generated from NS of the C4~O4~D14 model with 64 atoms. The yellow regions represent BCC structures and the blue regions represent HCP and FCC structures. \textbf{a} shows the region around the HCP-BCC transition and \textbf{b} shows the region around the BCC-FCC transition. Also shown are the predicted melting points (\textbf{a} and \textbf{b} cyan dots), the QHA for the HCP-BCC transition (\textbf{a} cyan plus symbols) and the BCC-FCC transition (\textbf{b} cyan triangles). Shown are clear solid-solid transitions, reflecting the expected boundaries. Note the pressure range between 34 and 50~GPa was sampled using replica-exchange-NS.\cite{unglert_replica_2025}}
\label{fig:qw_phase_dig}
\end{figure}

Due to the difficulty of resolving almost vertical solid-solid transition lines in \gls{ns}, we used the \gls{qha} to estimate the \gls{hcp}-\gls{bcc} and \gls{bcc}-\gls{fcc} phase boundaries. Due to the low computational requirements of the \gls{qha}, we fitted a higher order potential, C5~O4~D18, after performing an additional cycle and display these results as well.
Additionally, while the heat capacity peak is too shallow to pinpoint the exact temperature locations of any solid-solid transitions, they are clearly observed when looking at the order parameters shown in Figure~\ref{fig:qw_phase_dig}.

\Gls{ns} correctly samples the expected crystal structures across the entire pressure range, correctly predicting the transition from \gls{hcp} to \gls{bcc} to \gls{fcc}, without being explicitly provided with these structures, and without any external bias.
No prior assumptions on the phases were applied in constructing our database or producing the validation results.
The \gls{qha} for the \gls{hcp}-\gls{bcc} solid-solid transition agrees well with \gls{dft} based \gls{qha} results from Moriarty \textit{et al.}\cite{moriarty_first-principles_1995} and, as discussed by Moriarty \textit{et al.}, the disagreement with experiment stems from entropic effects stabilising \gls{bcc} at high temperature due to the soft phonon mode seen in Figure~\ref{fig:phonons}.
At low temperature stabilisation is caused by electronic effects aligning well with the \gls{qha}.
Though we are not able to fully reproduce the experimental data, we attribute this to the choice of technique rather than limitations of the model.

For the \gls{bcc}-\gls{fcc} solid-solid transition line our results show excellent agreement with the \textit{ab initio} \gls{qha} results produced by Li \textit{et al.}\cite{li_crystal_2010} and they agree similarly well to that of Smirnov.\cite{smirnov_comparative_2024} 
The boundary is predicted to have a positive, albeit very steep, gradient, but
due to the sensitivity of this phase boundary, the gradient value has a high uncertainty, including the possibility of taking negative values.

To complement the thermodynamic information with structural insight, we calculated the average Steinhardt $W_6$ parameter from our \gls{ns} simulations.
Heatmaps of the $W_6$ order parameter shown in Figure~\ref{fig:qw_phase_dig}, indicate a triple point between 480-500~GPa, hence there appears to be no inflection point at the triple point like the one suggested by Smirnov~\cite{smirnov_comparative_2024} or seen in lithium,\cite{frost_high-pressure_2019} and this \gls{bcc}-\gls{fcc} transition more closely resembles the character of that seen in sodium and potassium.\cite{polsin_structural_2022,narygina_melting_2011,mcbride_one-dimensional_2015}

The position of transitions obtained from the \gls{qha} align well with the $W_6$ order parameter plots but the fact that the 0~K enthalpy transition is at a higher pressure than the \gls{ns} 0~K enthalpy transition suggests the gradient of the line is negative, contrary to the \gls{qha}. 
Considering the steepness of the gradient of this boundary this prediction is within the uncertainty limitation of our approach, though further finite temperature studies are needed to confirm the nature of the boundary.

\subsection{Defects}
Through the investigation of defects in magnesium, we show that our database produced automatically from thermodynamic equilibrium structures, is extensive enough to accurately model vacancy formation enthalpies, self-interstitial formation enthalpies, and stacking faults, across our sampled pressure range.
In doing so, we demonstrate our \gls{mlip} can make valuable predictions under non-equilibrium conditions, even though defect configurations were not explicitly included in the training data.

The vacancy formation enthalpies for \gls{hcp}, \gls{bcc}, and \gls{fcc} under 0, 100, 460, and 500~GPa are given in Table~\ref{tab:vac_form_res} along with the minimum interatomic distance in the relaxed defect structures.
Across the broad pressure range of 0 to 500~GPa, and across the three structures, the largest disagreements between our C5~O4~D18 model and \gls{dft} are around 0.2~eV and 0.02~\AA.

\begin{table*}
\centering
\caption{\textbf{Vacancy formation enthalpies, $\boldsymbol{H_{vac}}$, and minimum interatomic distances, $\boldsymbol{r_{min}}$, for HCP, BCC, and FCC, at 0, 100, 460, and 500~GPa.} Values were calculated using DFT and using our C5~O4~D18 ACE model. $H_{vac}$ is in eV and $r_{min}$ is in \AA.}
\begin{ruledtabular}
\begin{tabular}{ cc|cc|cc|cc|cc }
 \multirow{2}{*}{Structure} & \multirow{2}{*}{Model} & \multicolumn{2}{c|}{0~GPa} & \multicolumn{2}{c|}{100~GPa} & \multicolumn{2}{c|}{460~GPa} & \multicolumn{2}{c}{500~GPa} \\
& & $H_{vac}$ & $r_{min}$ & $H_{vac}$ & $r_{min}$ & $H_{vac}$ & $r_{min}$ & $H_{vac}$ & $r_{min}$ \\
\hline
 \multirow{2}{*}{HCP (47 atoms)} & DFT & 0.79 & 3.15 & 4.69 & 2.42 & -- & -- & -- & -- \\
                           & C5~O4~D18 & 0.67 & 3.14 & 4.81 & 2.42 & -- & -- & -- & -- \\
\hline
 \multirow{2}{*}{BCC (35 atoms)} & DFT & 0.49 & 3.05 & 3.95 & 2.40 & 8.41 & 1.98 & 8.22 & 1.89 \\
                           & C5~O4~D18 & 0.35 & 3.03 & 4.14 & 2.40 & 8.15 & 1.99 & 8.13 & 1.88 \\
\hline
 \multirow{2}{*}{FCC (35 atoms)} & DFT & -- & -- & -- & -- & 9.10 & 1.95 & 9.41 & 1.92 \\
                           & C5~O4~D18 & -- & -- & -- & -- & 8.86 & 1.96 & 9.25 & 1.93 \\
\end{tabular}
\end{ruledtabular}
\label{tab:vac_form_res}
\end{table*}

In Table~\ref{tab:sf_form_res} we present the stacking fault formation energies, under 0~GPa, for four key structures that we identified in Section~\ref{sec:xrd_pats}. 
As above, these non-equilibrium formation energies show excellent agreement with \gls{dft}, with disagreements on the order of 0.005~eV.

\begin{table}
\centering
\caption{\textbf{Stacking fault formation energies from HCP at 0~GPa for four key stacking variants.} Results were calculated using DFT and our C5~O4~D18 ACE model. We found these four structures to express the unassigned XRD peaks in Figure~\ref{fig:layer_xrds}. We also provide the difference in predicted potential energies for the structures to highlight the close agreement.}
\begin{ruledtabular}
\begin{tabular}{ c|ccc }
\multirow{2}{*}{Structure} & \multicolumn{3}{c}{Formation energy, [eV]} \\
                            & DFT & C5~O4~D18 & ACE-DFT \\
\hline
ABAC ABAB ABAB & 0.022 & 0.020 & -0.00165 \\
ABAC BCBC BCAB & 0.036 & 0.035 & -0.00083 \\
ABAC BCBC BCBC & 0.042 & 0.037 & -0.00423 \\
ABAC ACAC BCAB & 0.042 & 0.037 & -0.00402 \\
\end{tabular}
\end{ruledtabular}
\label{tab:sf_form_res}
\end{table}

High energy non-equilibrium properties are typically the most challenging to compute with \glspl{mlip} on account of small interatomic distances which often easily leads into \gls{pes} holes.
However, in Table~\ref{tab:sie_res} we again show excellent agreement with \gls{dft} across the wide pressure range and across the relevant structures.
The general trends, such as the increasing enthalpy with increasing pressure, and the decrease between 460 and 500~GPa for the \gls{bcc} structures, are in complete agreement with the \gls{dft} results.
Additionally, the difference in interstitial enthalpy predictions varies from around 0.04~eV for the 0~GPa predictions up to around 0.5~eV for the highest pressure enthalpies, which is an excellent agreement across such a broad pressure range.
The minimum interatomic distances are also almost exact with the largest differences on the order of 0.01~\AA. The exception is the low pressure \gls{bcc} tetrahedral interstitial defect, due to \gls{bcc} being unstable at low pressures and therefore poorly sampled, the interstitial interatomic difference is 0.07~\AA.

\begin{table*}
\centering
\caption{\textbf{Self-interstitial enthalpies, $\boldsymbol{H_{si}}$, and minimum interatomic distances, $\boldsymbol{r_{min}}$, for HCP and BCC at 0,100,460, and 500~GPa.} For HCP the interstitial atom was placed in the tetrahedral and octahedral sites below the basal plane (B$_T$ and B$_O$ respectively), and for BCC, in the tetrahedral and octahedral sites ($T$ and $O$ respectively). Values were calculated using DFT and our C5~O4~D18 ACE model. $H_{si}$ is in eV and $r_{min}$ is in \AA. }
\begin{ruledtabular}
\begin{tabular}{ ccc|cc|cc|cc|cc }
\multirow{2}{*}{Structure} & \multirow{2}{*}{Site} & \multirow{2}{*}{Model} & \multicolumn{2}{c|}{0~GPa} & \multicolumn{2}{c|}{100~GPa} & \multicolumn{2}{c|}{460~GPa} & \multicolumn{2}{c}{500~GPa}\\
& & & $H_{si}$ & $r_{min}$  & $H_{si}$ & $r_{min}$ & $H_{si}$ & $r_{min}$ & $H_{si}$ & $r_{min}$ \\
\hline
\multirow{4}{*}{HCP (37 atoms)} &\multirow{2}{*}{B$_O$} & DFT & 2.89 & 2.47 & 6.66 & 1.92 & -- & -- & -- & -- \\
                                &                 & C5~O4~D18 & 2.94 & 2.47 & 6.50 & 1.92 & -- & -- & -- & -- \\
                                &\multirow{2}{*}{B$_T$} & DFT & 3.34 & 2.54 & 7.68 & 1.95 & -- & -- & -- & -- \\
                                &                 & C5~O4~D18 & 3.38 & 2.53 & 7.50 & 1.95 & -- & -- & -- & -- \\
\hline
\multirow{4}{*}{BCC (55 atoms)} &\multirow{2}{*}{$O$} & DFT & 2.48 & 2.58 & 6.06 & 1.98 & 6.09 & 1.67 & 5.88 & 1.66 \\
                                &               & C5~O4~D18 & 2.42 & 2.58 & 5.96 & 1.98 & 6.54 & 1.67 & 6.38 & 1.66 \\
                                &\multirow{2}{*}{$T$} & DFT & 2.00 & 2.72 & 5.48 & 2.08 & 5.77 & 1.74 & 5.59 & 1.72 \\
                                &               & C5~O4~D18 & 1.93 & 2.65 & 5.38 & 2.08 & 6.07 & 1.75 & 5.91 & 1.73 \\
\end{tabular}
\end{ruledtabular}
\label{tab:sie_res}
\end{table*}

We find these benchmarks remarkable since our database consists of only samples automatically produced as a function of thermodynamic relevance.

\subsection{X-Ray Diffraction Patterns} \label{sec:xrd_pats}
\begin{figure*}
\begin{center}
\includegraphics[width=\linewidth]{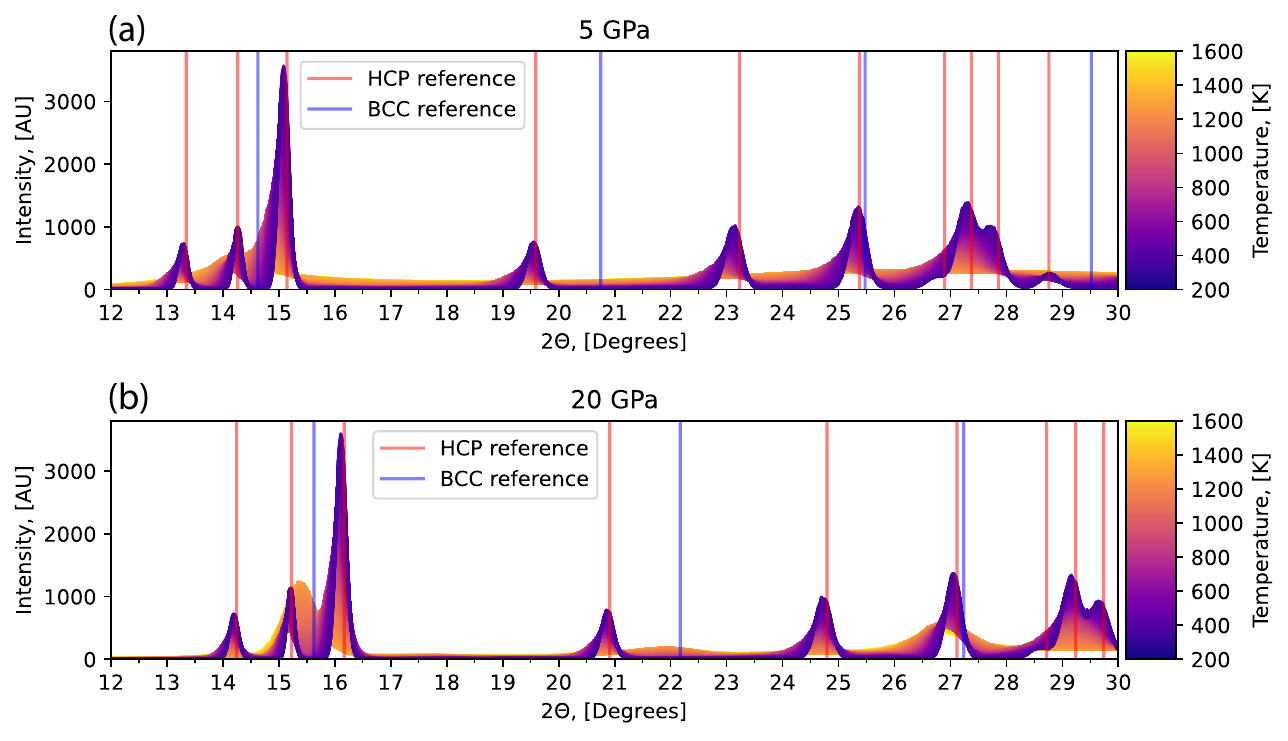}
\end{center}
\vspace{-20pt}
\caption{\textbf{XRD powder spectra calculated at different temperatures and pressures.} Using the XRD patterns of the samples collected from sampling the C4~O4~D14 model with 64 atoms at 5~GPa (panel~\textbf{a}) and 20~GPa (panel~\textbf{b}), we show the XRD patterns (coloured by temperature), against peaks for the enthalpy minimised BCC structure (blue) and HCP structure (red).}
\label{fig:xrd_pat}
\end{figure*}

The final property we investigated was the temperature dependent \gls{xrd} patterns, in order to interpret the additional peaks recorded by Stinton \textit{et al.} at a $2\theta$ angle of $15.9^\circ$ and $25.6^\circ$.\cite{stinton_equation_2014}
We did not observe any indication of these peaks, indicating that it is unlikely that these peaks correspond to an unidentified stable phase.
It follows that if they correspond to a single crystalline phase, it could only correspond to a metastable form.
In Figure~\ref{fig:xrd_pat} we present the temperature dependent \gls{xrd} patterns at 5 and 20~GPa to show that only the \gls{hcp} and \gls{bcc} phases are sampled in a substantial amount.

\begin{figure*}
\begin{center}
\includegraphics[width=\linewidth]{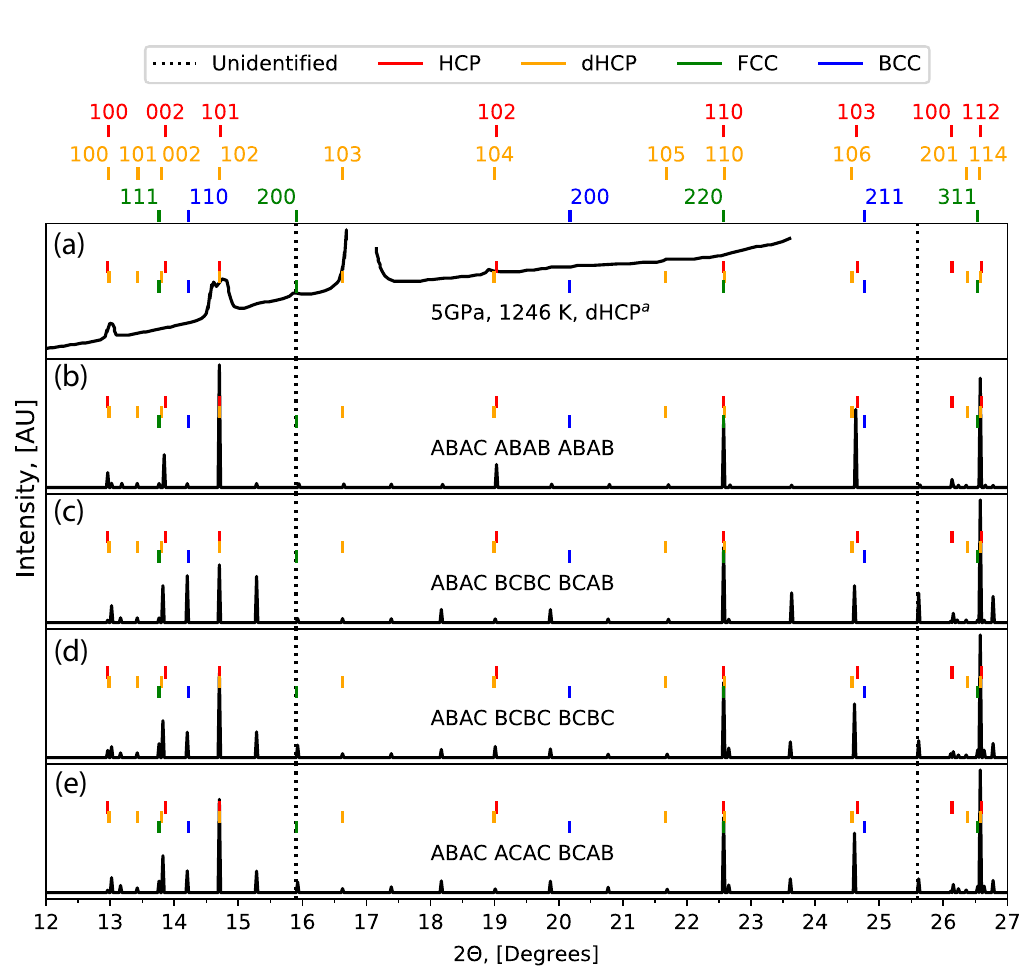}
\end{center}
\vspace{-20pt}
\caption{\textbf{Comparison of calculated XRD patters to experimental measurements.} We present the XRD patterns for four 12-layered close-packed structures (panels \textbf{b-e}), which all have the experimentally observed unassigned peaks, at a $2\theta$ of 15.9$^\circ$ and 25.6$^\circ$ (black dotted lines), reported by Stinton \textit{et al.}\cite{stinton_equation_2014} (panel \textbf{a}). Reference peaks are given from the enthalpy minimised HCP (red), dHCP (orange), FCC (green), and BCC (blue) structures, with the peak labels given in the top row.}
\label{fig:layer_xrds}
\end{figure*}

While studying the configurations sampled at 1~GPa and high temperatures, we observe a range of close-packed polytype structures.
These polytype structures included amounts of \gls{dhcp} (ABAC) and \gls{fcc} (ABC).
Further \gls{ns} runs with 21 atoms allowed us to also sample the 9R structure (ABACACBCB).
While none of these structures are predicted to have an \gls{xrd} peak at $25.6^\circ$, \gls{fcc} shows a peak at $15.9^\circ$.
Based on this observation, we systematically generated close-packed structures with up to 12 layers to identify if a long-period stacking order was responsible for the unidentified peaks.
We found multiple 12 layer structures, shown in Figure~\ref{fig:layer_xrds}, which show peaks at all positions reported experimentally.

\subsection{Computational Expense}
It is important to acknowledge the computational cost of carrying out our suggested procedure, as it has significant implications in terms of energy use and carbon footprint of high-performance computing.\cite{jackson_emissions_2023}
We discuss this cost in terms of CPUhrs, where 1~CPUhr is running 1 core for 1 hour.
We have neglected the computational cost of building the databases and training the linear \gls{ace} models as these are negligible compared to the \gls{dft} and \gls{ns} simulations.
The cost of the \gls{dft} per typical atomic configurations is primarily determined by the composition of the targeted system and would be similar in other comparable \gls{mlip} workflows.
In general, we aimed for a convergence resulting in sub \si{\meV\per atom} accuracy in our \gls{dft} calculations guiding our choice of parameters.
All calculations were performed on nodes of 2 AMD EPYC 7742 (Rome) 2.25 GHz 64-core processors.
Performing 16-atom \gls{dft} with our chosen parameters resulted in a cost of 70~CPUhrs/configuration and with 4330 16-atom configurations in total, the cost amounted to 303.1K~CPUhrs.
Performing 8-atom \gls{dft} with our parameters resulted in a cost of 13.5~CPUhrs/configuration and for 3900 configurations this amounts to a total of 52.7K~CPUhrs.
In total, around 346K CPUhrs of computational time were used for the \gls{dft} calculations of our training data points.

For the \gls{ns} component, the cost depends on the number of walkers, the walk length, and the model. In turn the model determines the cost with regard to the number of atoms and the pressure.
Within the \gls{ace} framework, the spatial cut-off of the \gls{mlip} is fixed so
at higher pressures, due to the decreased volume per atom, there is a significant increase in the number of atoms that fall within the cut-off of the potential, increasing the length of neighbour lists and evaluation times, leading to a marked rise in computational expense. Table~\ref{tab:cpuhrs} shows the cheapest and most expensive \gls{ns} runs with varying the model, the number of atoms, the number of walkers, and the pressure.\\

\begin{table}
\begin{threeparttable}
\centering
\caption{\textbf{CPUhrs taken for NS with different numbers of atoms at different pressures.} The walk length was the same in all cases.}
\begin{ruledtabular}
\begin{tabular}{ ccccc }
\multirow{2}{*}{Potential} & \multirow{2}{*}{Atoms} & Pressure & Cost     & Wall time\\
                           &                        & [GPa]    & [CPUhrs] & [hrs]    \\
\hline
EAM* & 16 & 0   & 408    & 8.5   \\
ACE  & 16 & 0   & 252    & 9.0   \\
ACE  & 64 & 1   & 3,144  & 131.0 \\
ACE  & 16 & 600 & 1,120  & 46.0  \\
ACE  & 64 & 600 & 11,648 & 485.0 \\
\end{tabular}
\end{ruledtabular}
\begin{tablenotes}
      \small
      \item *This was performed with four times as many walkers as with the \gls{ace}
\end{tablenotes}
\label{tab:cpuhrs}
\end{threeparttable}
\end{table}

\subsection{Summary}

In this study, we proposed and demonstrated a new framework for generating training data to develop \glspl{mlip}. Our approach stands out in its ability to automate the construction of a database based on thermodynamically relevant configurations rather than relying on human-driven selection of structures or pre-existing knowledge of certain phases. This ensures that the resulting \gls{mlip} captures phase properties reliably across a wide range of conditions.

Our procedure adapts an iterative training cycle to improve the performance of the \gls{mlip} model, with the \gls{ns} technique being at the heart of each cycle due to its ability to explore the relevant configuration space in an unbiased way. \Gls{ns} provides both critical information about the performance of the model and generates important configurations that need to be incorporated into the training dataset. This is particularly important for creating reliable \glspl{mlip} that are robust across volume, pressure, and temperature variables.

By applying this framework to magnesium as a test case, we successfully trained an \gls{mlip} using the \gls{ace} architecture, leveraging the committee \gls{std} of total energy predictions, to describe the material's behaviour over an extensive pressure-temperature range, covering liquid and solid phases up to 600~GPa.
We used an \gls{eam} model as a starting point, and the final potential was achieved after five training cycles, with the final database consisting of 4330 16-atom configurations and 3900 8-atom configurations.
We constructed two \gls{ace} models with: body order 4, degree 14 and a total of 710 basis functions; and a higher accuracy model with body order 4, degree 18 and a total of 2849 basis functions. Both potentials demonstrated excellent agreement with benchmark calculations, including geometry optimizations, phonon spectra, elastic constants, vacancy and self-interstitial formation enthalpies, and the phase diagram.

We used our potentials to probe the phase diagram at pressures which are challenging to achieve in experiments and to explore configuration space where experimental results are not completely explained.

In conclusion, the proposed framework represents a powerful and generalisable tool for developing \glspl{mlip}, with applications extending to a wide variety of materials and conditions. Its automated, thermodynamically informed, and extensible nature makes it a significant step toward overcoming current challenges in the field and enabling more accurate and efficient materials modelling.

\section{Methods}

\subsection{Nested sampling}
All \gls{ns} calculations were carried out using the \verb|pymatnest| software package.\cite{baldock_constant-pressure_2017, partay_nested_2021}
When generating and expanding the database, some of the \gls{ns} parameters were changed between active learning cycles.\\[12pt]
At each cycle the following parameters were kept constant. \Gls{ns} was performed starting from a maximum volume of 500~{\AA$^{3}$}/atom and progressing the sampling down to 200~K, culling one walker per iteration. The walk length for each walker retained the same ratio of moves given in Table~\ref{tab:ns_const_vars}.
A minimum allowed cell aspect ratio of 0.65 was used which was increased to 0.95 for the 64-atom \gls{ns}.\cite{baldock_classical_2017} When we restricted accepted walk moves to produce configurations where the \gls{std} of total energy predictions from the committee are below 62.5~meV/atom, we stopped the \gls{ns} run if 90\% of the walkers had a value above 60.625~meV/atom.
\begin{table}
\begin{threeparttable}
\centering
\caption{\textbf{MC moves used during NS.} Provided are the type and proportion of different MC moves used during NS walks and the acceptance rates used to dynamically adjust step sizes.}
\begin{ruledtabular}
\begin{tabular}{lcc}
Move Type & Proportion & Acceptance Rate\\
\hline
5-step TEHMC*  & 0.21 & 50-95\%\\
Volume Change & 0.31 & 25-75\%\\
Cell Shear    & 0.24 & 25-75\%\\
Cell Stretch  & 0.24 & 25-75\%\\
\end{tabular}
\end{ruledtabular}
\begin{tablenotes}
      \small
      \item *Total enthalpy Hamiltonian Monte Carlo~\cite{baldock_constant-pressure_2017}
\end{tablenotes}
\label{tab:ns_const_vars}
\end{threeparttable}
\end{table}

Across the cycles, the number of walkers, $K$, the sampled pressures, the walk length, and maximum committee \gls{std} was varied due to the associated cost, required accuracy, and stability of the different models, as discussed in detail in the previous sections. These parameters are given in Table~\ref{tab:ns_variables}.
\begin{table}
\centering
\caption{\textbf{NS parameters used across the active learning cycles.} The number of atoms, number of walkers, $K$, number of proposed steps in decorrelating the configurations between iterations, $L$, and the maximum accepted STD of energy predictions made by the committee of models (meV/atom).}
\begin{ruledtabular}
\begin{tabular}{cclccc}
\multirow{2}{*}{Cycle} & \multirow{2}{*}{Atoms} & \multicolumn{1}{c}{Pressure range} & \multirow{2}{*}{$K$} & \multirow{2}{*}{$L$} & Comm. STD\\
      &       & \multicolumn{1}{c}{[GPa]}          &     &     & [meV/atom] \\
\hline
0   & 16 & 0, 1, [5-45,5] & 1248 & 1248 & N/A \\
1   & 16 & 0, 1, [5-45,5], [60-600,20] & 336  & 336  & 62.5 \\
2   & 16 & 0, 1, [5-45,5], [60-600,20] & 336  & 336  & 62.5 \\
3   & 16 & 0, 1, [5-45,5], [60-600,20] & 336  & 336  & None \\
4   & 64 & 0, 1, [5-45,5], 100,200,300 & 336  & 336  & 62.5 \\
4   & 64 & [60-90,10], [400-600,20] & 384  & 336  & None \\
\end{tabular}
\end{ruledtabular}
\label{tab:ns_variables}
\end{table}

\subsection{ACE models}
All \gls{ace} fitting was done using the \verb|ACEsuit| Julia software package.\cite{drautz_atomic_2019, dusson_atomic_2022}
The cutoff for constructing the \gls{ace} was set to 8.2~{\AA}, as at this distance the \gls{dft} pair potential is less than 10$^{-8}$~eV/atom.  This value has also been used in related studies of magnesium.\cite{ibrahim_atomic_2023} A single atom reference energy of $-1688.821$~eV was used, calculated by placing a single Mg atom in a suitably large cubic cell.

In order to pick ideal parameters for the \gls{ace} model, once the initial database was constructed, 20\% of the configurations were randomly selected and removed from the training set, forming a test set.
The model was then trained on the remaining 80\% of the data and the accuracy of the fit was determined by calculating the \gls{rmse} of the predicted energies of the test set.
During the initial fitting, the configurations were weighted equally, and the energies, forces, and virial stress components were weighted with a ratio of 9:1:1.
The chosen body order was set at 4 and the degree was set at 14, for a potential consisting of 710 basis functions.
Once these parameters were determined, a refit was performed using the full database.
The \gls{rmse} for the fit of this potential to the training data was 1.3~meV, 19~meV/\AA, 10.7~meV, for the total energy, forces, and virials respectively.
Bayesian linear regression was employed for the fitting and from the produced posterior distribution, ten parameter sets were drawn, forming a committee of potentials.\cite{witt_acepotentialsjl_2023}
This was used to evaluate the uncertainty associated with energy predictions made by the \gls{ace} model, by calculating the \gls{std} of the ten total-energy predictions made by the committee.\\

In our active learning procedure, we used Equation~\ref{eqn:reweight} to calculate the individual weight, $W_i$, for each configuration in the loss function, using the enthalpy difference between the configuration generated at the $i$-th iteration, $H_i$, and the enthalpy of the final sample, $H_F$ generated during a \gls{ns} run. We controlled this exponential through parameter $\alpha$ and found a value of $0.1$ minimised the \gls{rmse} during fitting.

\begin{equation} \label{eqn:reweight}
\begin{aligned}
    W_i &= e^{-\alpha (H_i - H_F)} \\
\end{aligned}
\end{equation}

\subsection{DFT}\label{sec:dft}
All configurations within the database were evaluated with the \verb|CASTEP| \gls{dft} software package,\cite{clark_first_2005} using the \gls{pbe} exchange correlation functional.\cite{perdew_generalized_1996}
Mg was represented by an on-the-fly generated ultra-soft pseudopotential based on the C19 definition in \verb|CASTEP|, with a core radius of 1.8~Bohr and 10 valence electrons explicitly considered in the configuration [2s2 2p6, 3s2].
A plane wave cutoff of 700~eV was used with a fine grid scale of 4.0 and an \gls{scf} convergence tolerance of 10$^{-9}$~eV.
\Gls{mp} k-point grids, with a maximal grid spacing of 0.015~{\AA$^{-1}$}, were used to sample the Brillouin zone and we applied Gaussian smearing to the occupancies with a width of 0.2~eV to improve convergence. Convergence tests and support for these \gls{dft} parameters can be found in Appendix~\ref{sec:dft_con_test}.

\subsection{Phonons}
The \gls{dft} phonon spectra were calculated using the finite displacement method implemented in the \verb|CASTEP| software package utilising non-diagonal supercells.\cite{clark_first_2005,lloyd-williams_lattice_2015}
In addition to the \gls{dft} parameters specified in Section~\ref{sec:dft}, a finite displacement of 0.05~{\AA} was used on minimum enthalpy structures produced using the parameters specified in Section~\ref{sec:enthalpy_min}.
A q-grid of $4\times4\times4$ was used and interpolated to a finer grid with maximal grid spacing of 0.1~{\AA$^{-1}$} along the high-symmetry paths to produce the \gls{dft} phonon spectra.

The \gls{ace} phonon spectra were calculated using the \verb|phonopy| Python software package.\cite{togo_implementation_2023, togo_first-principles_2023} 
Supercells of size $4\times4\times4$ were constructed for the four key crystal structures (\gls{hcp}, \gls{dhcp}, \gls{fcc}, and \gls{bcc}) and finite displacements of 0.05~{\AA} were used to determine the force constant matrices.

\subsection{Elastic Constants}
Both the \gls{ace} and \gls{dft} elastic constants were calculated using the \verb|matscipy| Python software package.\cite{grigorev_matscipy_2024}
In both cases the unit cells were first relaxed, and the finite strains were applied in increments of $5\times10^{-5}$.
The increments were chosen such that decreasing the finite strains further resulted in no significant change in the elastic constants.
The \gls{dft} parameters given in Section~\ref{sec:dft} were used, except all \gls{dft} grids were fixed to those in Table~\ref{tab:mp_grids}.

\begin{table}
\centering
\caption{\textbf{Fixed MP k-point grids used for elastic constant and enthalpy minimisation calculations for the different crystal structures unit cells.} These were chosen as these are the grids generated from a 0.015~{\AA$^{-1}$} maximum grid spacing for the minimum enthalpy unit cell structures at 600~GPa.}
\begin{tabular}{cc}
\hline\hline
Crystal structure & MP k-point grid\\
\hline
HCP   & $38\times38\times20$ \\
dHCP  & $38\times38\times10$ \\
BCC   & $42\times42\times42$ \\
FCC   & $41\times41\times41$ \\
\hline\hline
\end{tabular}
\label{tab:mp_grids}
\end{table}

\subsection{Enthalpy Minimisations and Defects}\label{sec:enthalpy_min}
All \gls{dft} calculations were performed using \verb|CASTEP| with the parameters specified in Section~\ref{sec:dft}, unless otherwise specified below.\cite{clark_first_2005}
Tolerances of 0.02~meV, 1~meV/\AA, 0.01~GPa, and 0.001~\AA~were used to determine convergence for the total energy, forces, stresses, and atomic displacements respectively for any \gls{dft} geometry optimisations.
Equivalent calculations using a \gls{mlip} were performed using the \verb|Atomic Simulation Enviornment (ASE)| software package, with a force tolerance of $10^{-5}$~eV/\AA~to determine convergence.\cite{hjorth_larsen_atomic_2017}

For calculating the enthalpy minima, the symmetry of the unit cells was maintained by fixing the angles and ratios of relevant cell parameters during optimisation, and to avoid k-point re-meshing, fixed k-point grids, given in Table~\ref{tab:mp_grids}, were used throughout the minimisations.

To calculate the vacancy formation enthalpies, we used the minimum enthalpy structures under the pressures specified in Table~\ref{tab:vac_form_res}, to construct supercells of \gls{hcp}, \gls{bcc}, and \gls{fcc}, and a single atom removed. With the lattice parameters fixed, the structures were relaxed to the tolerances specified above. The formation enthalpy was then calculated using Equation~\ref{eqn:vac_form_enth}.

\begin{equation} \label{eqn:vac_form_enth}
\begin{aligned}
    H_{vac} =& \big( U_{vac}(N-1) - \frac{N-1}{N}U_{bulk}(N)\big) \\
            &+ P\big( V_{vac}(N-1) - \frac{N-1}{N}V_{bulk}(N)\big)\\
\end{aligned}
\end{equation}

To calculate the self interstitial enthalpies, we used the minimum enthalpy structures under the given pressures to construct supercells of \gls{hcp} and \gls{bcc}. 
For \gls{hcp} a single atom was added to the $B_O$ and $B_T$ positions, and for \gls{bcc} a single atom was added to the $O$ and $T$ sites.
With the lattice parameters fixed, the structures were relaxed, and the self-interstitial enthalpy was calculated using Equation~\ref{eqn:sie_enth}.

\begin{equation} \label{eqn:sie_enth}
\begin{aligned}
    H_{si} =& \big( U_{si}(N+1) - \frac{N+1}{N}U_{bulk}(N)\big) \\
            &+ P\big( V_{si}(N-1) - \frac{N+1}{N}V_{bulk}(N)\big)\\
\end{aligned}
\end{equation}

Finally to calculate the stacking fault formation energies, we used the lattice parameters from the 0~GPa minimum enthalpy \gls{hcp} structure to construct the four stacking variants provided in Table~\ref{tab:sf_form_res}.
With the lattice parameters fixed, the structures were relaxed, and the formation energies calculated by subtracting the potential energy of an equally sized pure \gls{hcp} bulk.

\subsection{Bain Path}
Starting from the \gls{bcc} unit cell enthalpy minimum at 460~GPa, the unit cell was elongated in the $c$ direction and then relaxed with a fixed $c/a$ ratio at 460~GPa. This was done for ten $c/a$ ratios from 1 to $\sqrt{2}$ to give the lowest enthalpy pathway along the Bain path.
This was done using \gls{dft} and the parameters provided in Section~\ref{sec:dft} with \verb|CASTEP|, and with our \gls{ace} potential using \verb|ASE|.

\subsection{Stacking Variant Investigation}

To generate the stacking variants systematically, 1 to 12 atoms were equally placed along $c$ in a hexagonal cell ($a$,$a$,$c$,~$90^\circ$,$90^\circ$,$120^\circ$), with fractional $x$-$y$ positions of (0,0), (2/3,1/3), or (1/3,2/3). This results in 797,160 possible structures, without correcting for symmetry.
After removing equivalent structures, there are 11,676 unique stacking variants, with the quantity of the different layered structures given in Table~\ref{tab:stack_var_num}.
These configurations were relaxed under 1~GPa and the \gls{xrd} patterns calculated for a simulated x-ray wavelength of 0.62~\AA, using the \verb|QUIP| software package.\cite{csanyi_expressive_2007}
\begin{table}
\centering
\caption{\textbf{Number of possible unique stacking variants.} After accounting for symmetry, the number of unique stacking variants with up to 12 layers of one species in a hexagonal cell is 11,676, the quantity of unique structures for each $N$ layered structure is provided.}
\begin{tabular}{cc}
\hline\hline
Layers & Unique stacking variants\\
\hline
1 & 1     \\
2 & 1     \\
3 & 2     \\
4 & 4     \\
5 & 8     \\
6 & 22    \\
7 & 52    \\
8 & 140   \\
9 & 366   \\
10 & 992  \\
11 & 2684 \\
12 & 7404 \\
\hline\hline
\end{tabular}
\label{tab:stack_var_num}
\end{table}
\section*{Data availability}

The data supporting this study's findings can be found on GitHub at 
\url{https://github.com/VGFletcher/Fletcher_2025_magnesium_data}.

\section*{Code availability}

The open-source package \verb|pymatnest| is freely available on GitHub at \url{https://github.com/libAtoms/pymatnest}.
Python functions to construct databases from \verb|pymatnest| output files can be found at \url{https://github.com/VGFletcher/NS_database_builder}.


\begin{acknowledgments}

V.G.F. acknowledges funding from a studentship jointly by ICASE and AWE-Nuclear Security Technologies, and training support by the EPSRC Centre for Doctoral Training in Modelling of Heterogeneous Systems (EP/S022848/1).
A.B.P acknowledges support from the CASTEP-USER grant funded by UK Research and Innovation (EP/W030438/1).
L.B.P. acknowledges support from the EPSRC through the individual Early Career Fellowship (EP/T000163/1).
Computing facilities were provided by the Scientific Computing Research Technology Platform of the University of Warwick. 
Part of the calculations were performed using the Sulis Tier 2 HPC platform hosted by the Scientific Computing Research Technology Platform at the University of Warwick.
Sulis is funded by EPSRC Grant EP/T022108/1 and the HPC Midlands+ consortium.

\end{acknowledgments}

\section*{Author Contributions}
A.B.P and L.B.P designed the study, obtained funding and supervised the research.
V.G.F. developed the methodology, carried out the simulations and the analysis.
V.G.F prepared the figures for the manuscript.
All authors have contributed to writing and reviewing the manuscript text.


\section*{Competing Interests}
There are no conflicts to declare.

\bibliographystyle{naturemag}
\bibliography{main}
\appendix

\section{EAM Phase Diagram}\label{sec:EAM_pd}

\begin{figure}[H]
\begin{center}
\includegraphics[width=8.5cm,angle=0]{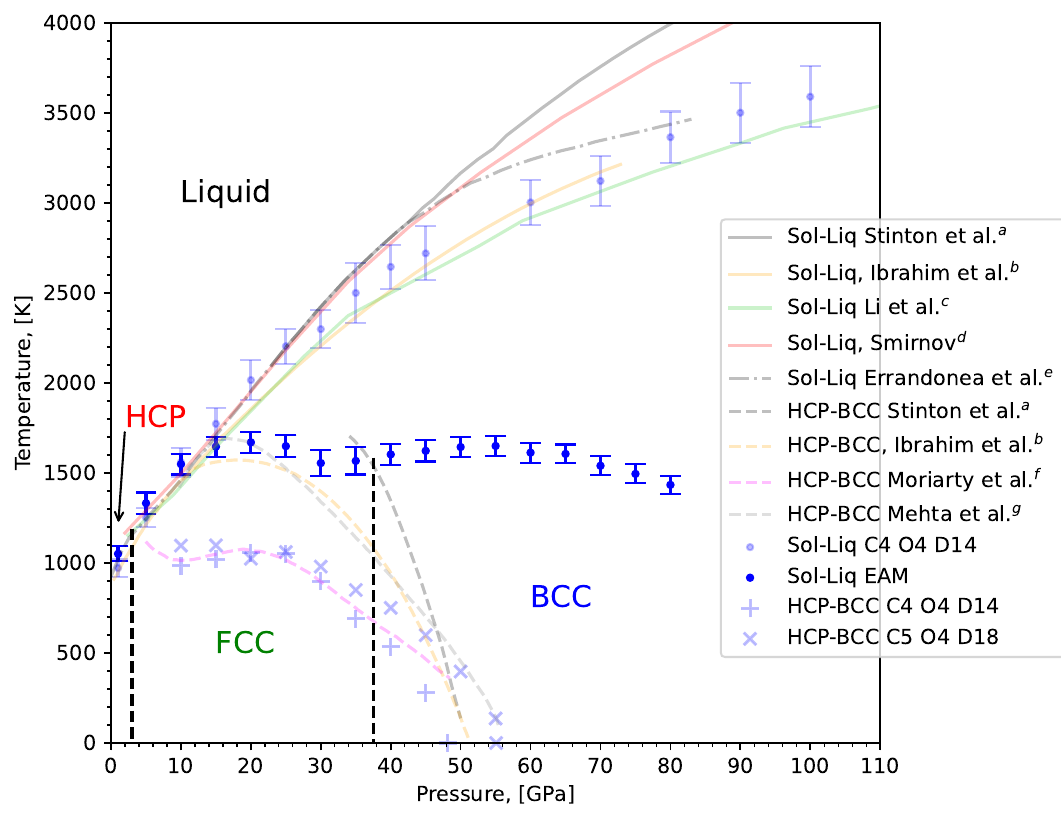}
\end{center}
\vspace{-20pt}
\caption {\textbf{Pressure-temperature phase diagram of magnesium for the Wilson \textit{et al.} EAM model.\cite{wilson_unified_2016}} Results from the EAM potential are shown in blue, and compared to predictions made using the C4~O4~D14 ACE potential developed in the current work (faded blue symbols) and to previous experimental measurements and computational predictions (a:\cite{stinton_equation_2014}, b:\cite{ibrahim_atomic_2023}, c:\cite{li_multiphase_2024}, d:\cite{smirnov_comparative_2024}, e:\cite{errandonea_melting_2001}, f:\cite{moriarty_first-principles_1995}, g:\cite{mehta_ab_2006}, h:\cite{li_crystal_2010}). The error bars on our NS results represent the full-width half-maximum of the calculated constant pressure heat capacity peaks (discussion on this can be found in Appendix~\ref{sec:error_bar}).}
\label{fig:eam_phase_diagram}
\end{figure}

The phase diagram of the magnesium \gls{eam} potential, produced by Wilson \textit{et al.},\cite{wilson_unified_2016} disagrees considerably with the \textit{ab initio} predictions and experimental observations, as shown in Figure~\ref{fig:eam_phase_diagram}. While the melting temperatures from 0-15~GPa agree reasonably with the expected results, there is an incorrect \gls{fcc} phase and a \gls{hcp} to \gls{fcc} solid-solid transition between 1 and 5~GPa. Additionally, past 40~GPa the correct stable phase of \gls{bcc} is predicted but the melting temperature is significantly different to that of the expected results and, due to the unrealistic \gls{fcc} phase, there is an incorrect \gls{fcc} to \gls{bcc} solid-solid transition between 35 and 40~GPa. We used this behaviour to produce our initial database.

\section{Error Bars in Nested Sampling}\label{sec:error_bar}

\Gls{ns} is carried out with a finite number of atoms (8-64 in this study), this results in finite size effects that are reflected through a peak on the temperature -- heat-capacity plots during phase transitions, rather than a discontinuity that would be seen in the macroscopic system.
It is observed that, as system size increases, these peaks become sharper and shift lower in temperature, with the shift becoming increasingly smaller as system size increases.\cite{marchant_exploring_2023}
Additionally, it is seen when repeating converged \gls{ns} runs, the position of the peak, which can shift due to the stochastic nature of the sampling, doesn't shift significantly regardless of system size.
Thus, to provide the most meaningful measure of the uncertainty of the position of a phase transition, we provide the \gls{fwhm} of the heat capacity peaks taken from a baseline positioned at the tail of the curve which produces the lowest peak prominence.

\section{Minimum Bond length Restriction}

Before we introduced the use of a committee, to control sampling of \gls{pes} holes, our first solution was to exclude configurations with unphysically short interatomic distances.
We were initially concerned with studying magnesium up to 100~GPa, and at this pressure there is a generous interatomic buffer zone to choose a minimum bond length that ensures only the very high-temperature configurations are effected by this restriction.
We were willing to allow this, since these configurations are not particularly important for our study.
However, when we expanded the pressure range of interest to up to 600~GPa, the parameter choice became difficult to choose such that it did not interfere with the sampling of the liquid phase. Thus we moved away from this solution.

\section{Bain Path}

\begin{figure}
\begin{center}
\includegraphics[width=8.5cm,angle=0]{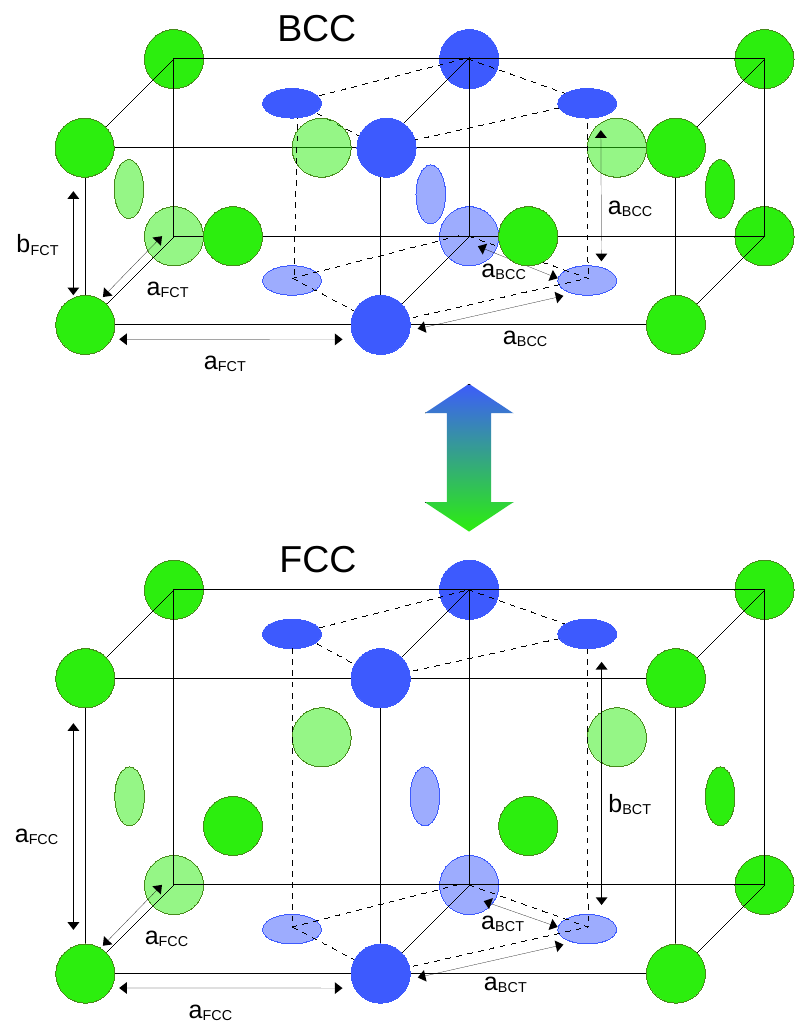}
\end{center} \vspace{-20pt}
\caption {\textbf{Schematic representation of the Bain transition pathway.} Represented is the transition from BCC to FCC through a BCT transition state, also known as the Bain path.}
\label{fig:bain_path} 
\end{figure}

A schematic for the lattice deformation from \gls{bcc} to \gls{fcc} through a \gls{bct} intermediate is displayed in Figure~\ref{fig:bain_path}.

\section{DFT Convergence Tests}\label{sec:dft_con_test}

\begin{figure*}
\begin{center}
\includegraphics[width=\linewidth,angle=0]{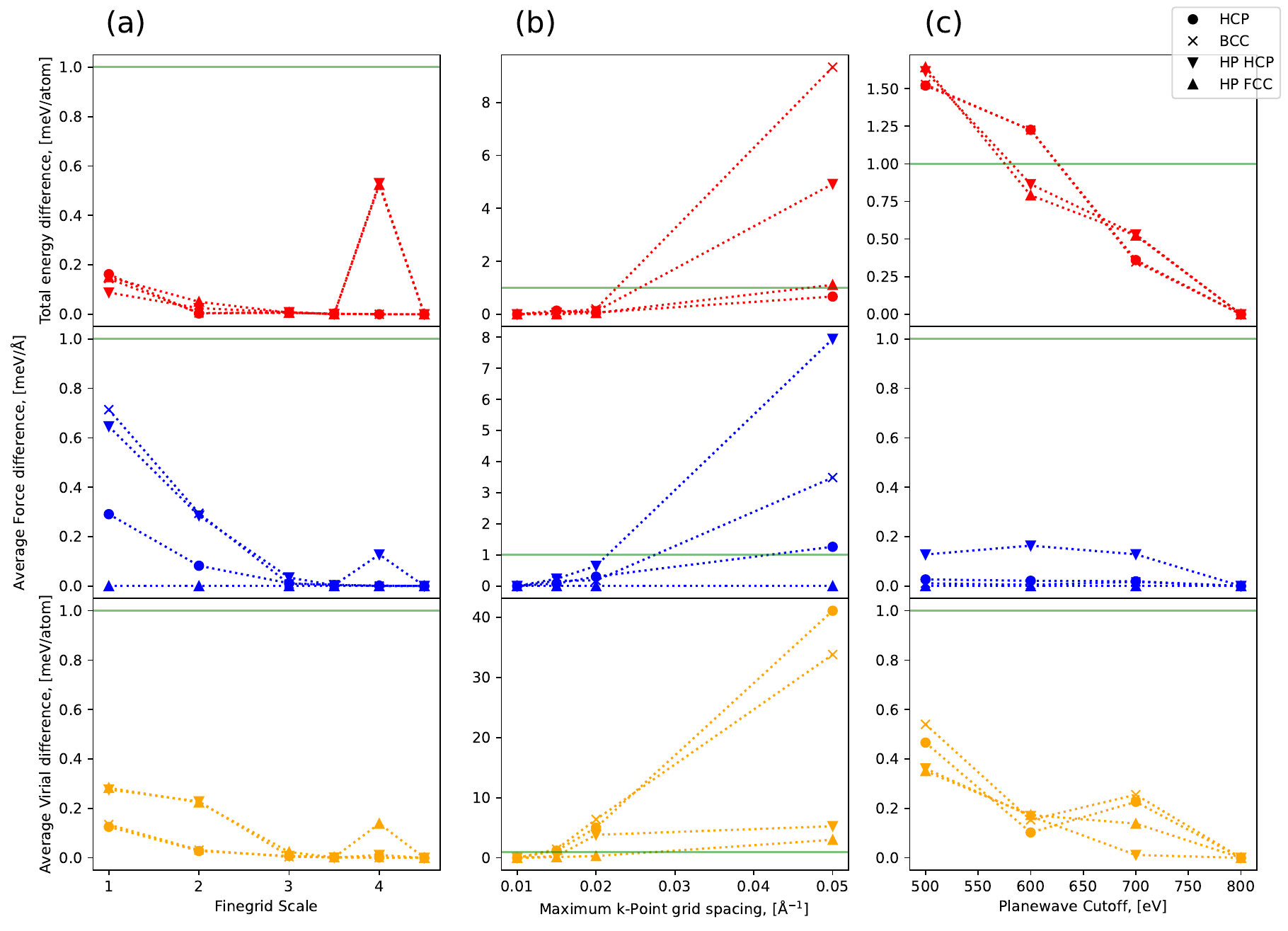}
\end{center}
\vspace{-20pt}
\caption {\textbf{Changes to total energy, forces, and virials, with respect to DFT parameters, for four key crystal structures.} Plots of the difference of total energy (red), the average forces (blue), and average virials (orange) to the most expensive parameter, with respect to finegrid scale (column \textbf{a}), maximum k-point grid spacing (column \textbf{b}), and planewave cutoff (column \textbf{c}). Results are given for supercells of a low and high density HCP structure (dots and down-pointing triangles respectively), a low density BCC structure (crosses), and a high density FCC structure (up-pointing triangles). Horizontal green lines show 1~meV/atom, below which we are satisfied convergence is reached.}
\label{fig:conv_test}
\end{figure*}

To find converged \gls{dft} parameters, $2\times2\times2$ supercells of the four unit cells given in Table~\ref{dft_conv_cells} were constructed. The cell volumes were increased by 5\%, the lattice vector components were perturbed randomly by 0-3\%, and the atomic positions randomly perturbed by $0-0.02$~\AA. 
The \gls{dft} parameters were chosen to achieve sub-meV/atom convergence with respect to the total energy, average sub-meV/atom with respect to components of the virial stresses, and average sub-meV/\AA convergence with respect to atomic forces. These results are shown in Figure~\ref{fig:conv_test}.

\begin{table}[H]
\centering
\caption{\textbf{Lattice parameters of the initial unit cells used for convergence tests.} Supercells of these unit cells underwent random perturbation, as described in Appendix~\ref{sec:dft_con_test}, to determine converged DFT parameters.}
\begin{tabular}{rccccccc}
\hline
\textbf{Crystal} & \textbf{a} (\AA) & \textbf{b} (\AA)& \textbf{c} (\AA)& \textbf{$\alpha$} ($^{\circ}$) & \textbf{$\beta$} ($^{\circ}$)& \textbf{$\gamma$} ($^{\circ}$) & \textbf{n atoms}\\
\hline
5~GPa HCP   & 3.079 & 3.079 & 4.99 & 90 & 90 & 120 & 2\\
5~GPa BCC   & 2.983 & 2.983 & 2.983 & 109.47 & 109.47 & 109.47 & 1\\
600~GPa HCP & 2.029 & 2.029 & 3.40 & 90 & 90 & 120 & 2\\
600~GPa FCC & 2.04 & 2.04 & 2.04 & 60 & 60 & 60 & 1\\
\hline
\end{tabular}
\label{dft_conv_cells}
\end{table}

\begin{figure}
\begin{center}
\includegraphics[width=8.5cm,angle=0]{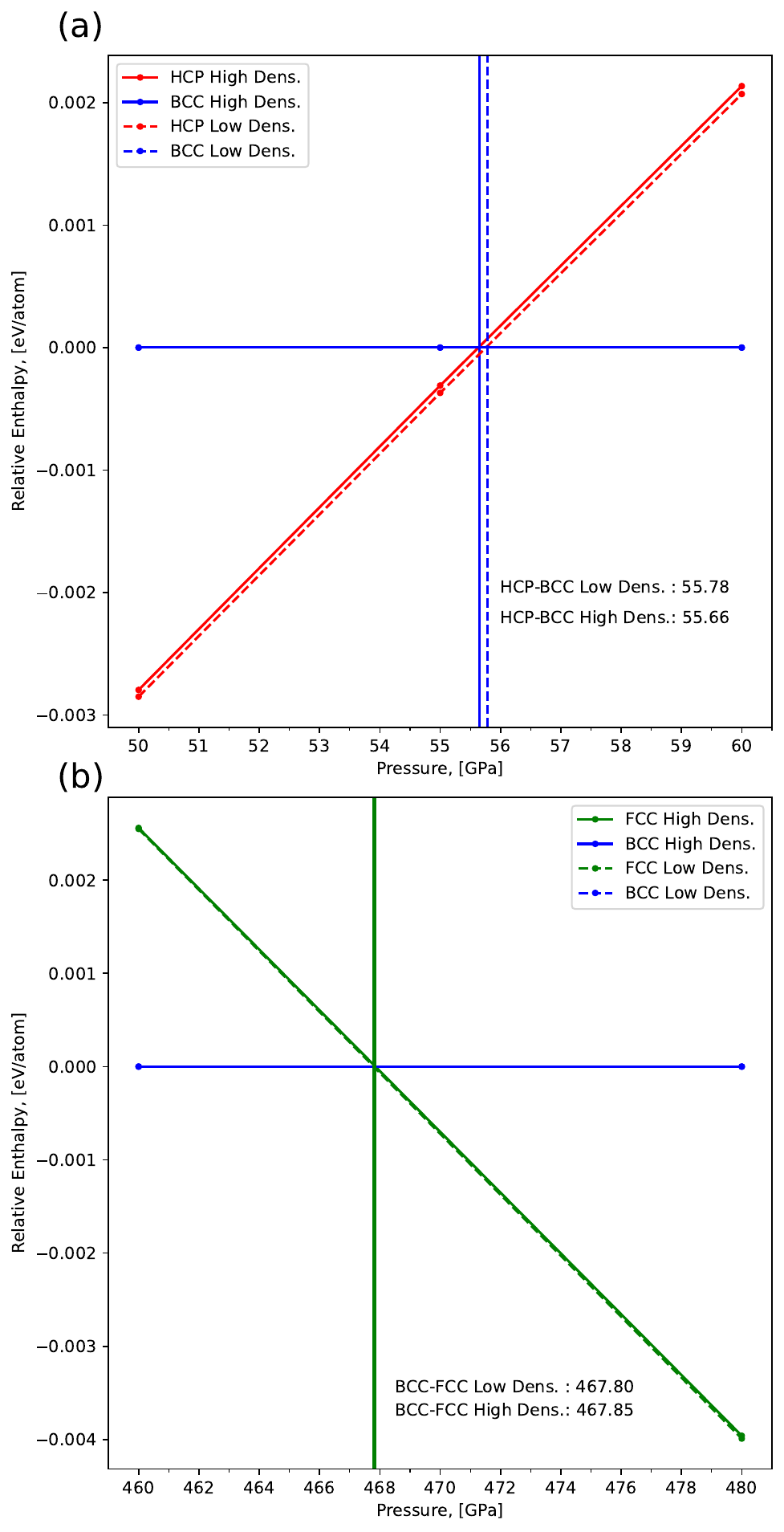}
\end{center}
\vspace{-20pt}
\caption {\textbf{The 0~K transition pressures for the two solid-solid transitions, calculated with different k-point grids.} \textbf{a} shows the HCP-BCC 0~K transition at around 55~GPa, while \textbf{b} shows the BCC-FCC 0~K transition at around 467~GPa. Increasing the grid density results in negligible changes to the transition pressures, supporting our choice of grid spacing. The grids are given in Table~\ref{mp_grids_fine}.}
\label{finer_grid}
\end{figure}

To verify the effect of our chosen convergence parameters, and show that our k-point convergence is acceptable, we increased the density of the \gls{mp} k-point grids, to those shown in Table~\ref{mp_grids_fine}, and recalculated the enthalpy minima. This resulted in a 0.12~GPa decrease in the \gls{hcp}-\gls{bcc} 0~K phase transition, and a 0.05~GPa increase in the \gls{bcc}-\gls{fcc} transition as shown in Figure~\ref{finer_grid}. These small changes support our choice of \gls{dft} parameters.

\begin{table}[H]
\centering
\caption{\textbf{MP k-point grids used during DFT enthalpy minimisations.} To show the effect on the 0~K phase transitions, we used the given sets of fixed k-point grids, to perform geometry optimisations, and calculate the transition pressures.}
\begin{tabular}{cc|c}
\hline
& \multicolumn{2}{c}{\textbf{MP k-point Grid}}\\
\hline
\textbf{Crystal} & \textbf{Original} & \textbf{Finer}\\
\hline
HCP   & $38\times38\times20$ & $39\times39\times21$\\
dHCP  & $38\times38\times10$ & $39\times39\times11$\\
BCC   & $42\times42\times42$ & $43\times43\times43$\\
FCC   & $41\times41\times41$ & $42\times42\times42$\\
\hline
\end{tabular}
\label{mp_grids_fine}
\end{table}

\section{Pseudopotential Delta Test}

To verify that the ultrasoft pseudopotential used in our \gls{dft} calculations was still accurate at 600~GPa, we calculated the delta gauge specified in the paper by Lejaeghere \textit{et al.} and given in Equation~\ref{delta}.\cite{lejaeghere_reproducibility_2016}

\begin{equation} \label{delta}
\Delta_i(a,b) = \sqrt{\frac{\int^{1.06V_{0,i}}_{0.94V_{0,i}}(E_{b,i}(V)-E_{a,i}(V))^2dV}{0.12V_{0,i}}}
\end{equation}

To do this, the \gls{fcc} unit cell - minimised at 600~GPa using the \gls{dft} parameters specified in the main paper - was scaled to produce the $E_{b,i}(V)$ curve.
Separately, the pseudopotential was changed to the hard pseudopotential specified in \verb|CASTEP|, and new \gls{dft} parameters were determined to achieve sub meV/atom total energy accuracy. This required a plane-wave cutoff of 1200~eV with all other parameters being acceptably converged for the change.
The \gls{fcc} unit cell was minimised again at 600~GPa using the new \gls{dft} parameters and this new minimum was scaled to produce the $E_{a,i}(V)$ curve, the key parameter differences are shown in Table~\ref{hard_psp_prop}. For simplicity, quadratic functions were fitted to the curves to allow easy calculation of the differences and integrals. The measured delta was 0.035 which is acceptably negligible to consider the two calculations in agreement.

\begin{table}[H]
\centering
\caption{\textbf{Key parameter changes when moving from the ultrasoft to the hard pseudopotential.} To prove convergence with respect to our choice of ultrasoft pseudopotential, we used a harder construction, and re-converged our DFT parameters. Shown are the differences in the planewave cutoff, needed to achieve convergence, and the change to the FCC lattice parameter.}
\begin{tabular}{ccc}
\hline
\textbf{Property} & \textbf{Ultrasoft} & \textbf{Hard}\\
\hline
Plane wave Cutoff, eV & 700       & 1200 \\
FCC a Lat. Par. \AA  &  2.042916 & 2.041489 \\
\hline
\end{tabular}
\label{hard_psp_prop}
\end{table}

\begin{figure}[H]
\begin{center}
\includegraphics[width=8.5cm,angle=0]{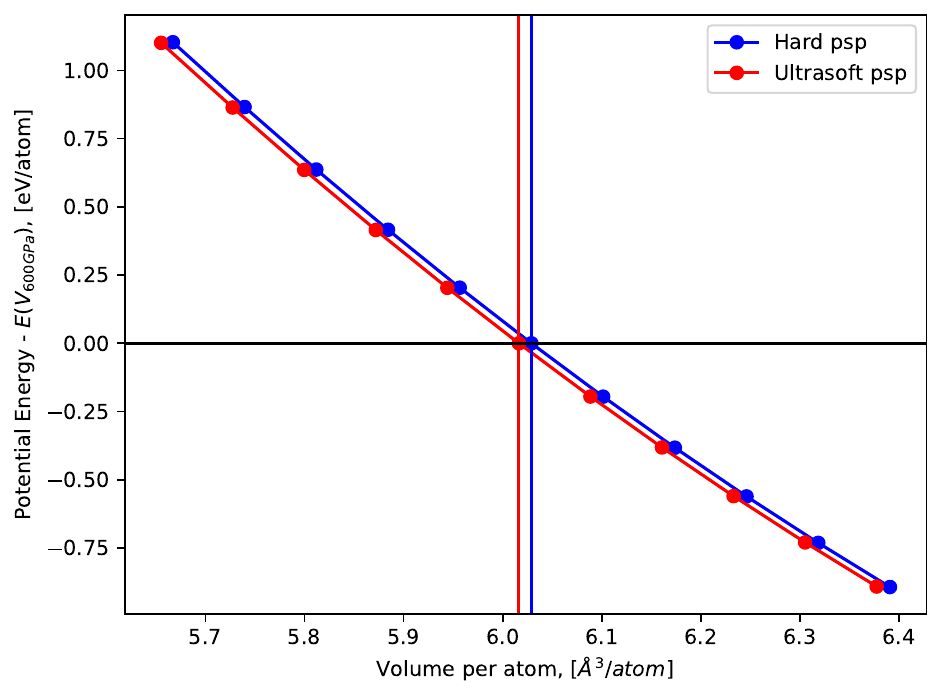}
\end{center}
\vspace{-20pt}
\caption {\textbf{Potential energy from isotropically scaling a unit cell of FCC, from the 600~GPa enthalpy minima of the different pseudopotentials.} Both instances used the enthalpy minimised structure for their respective parameter sets. The close agreement shows a negligible effect of changing to a more accurate pseudopotential at the smallest interatomic distances we investigated.}
\label{fig:delta_test}
\end{figure}

\section{Latent Heats and Enthalpy Curves}\label{sec:lh_and_enths}

\begin{figure}[H]
\begin{center}
\includegraphics[width=8.5cm,angle=0]{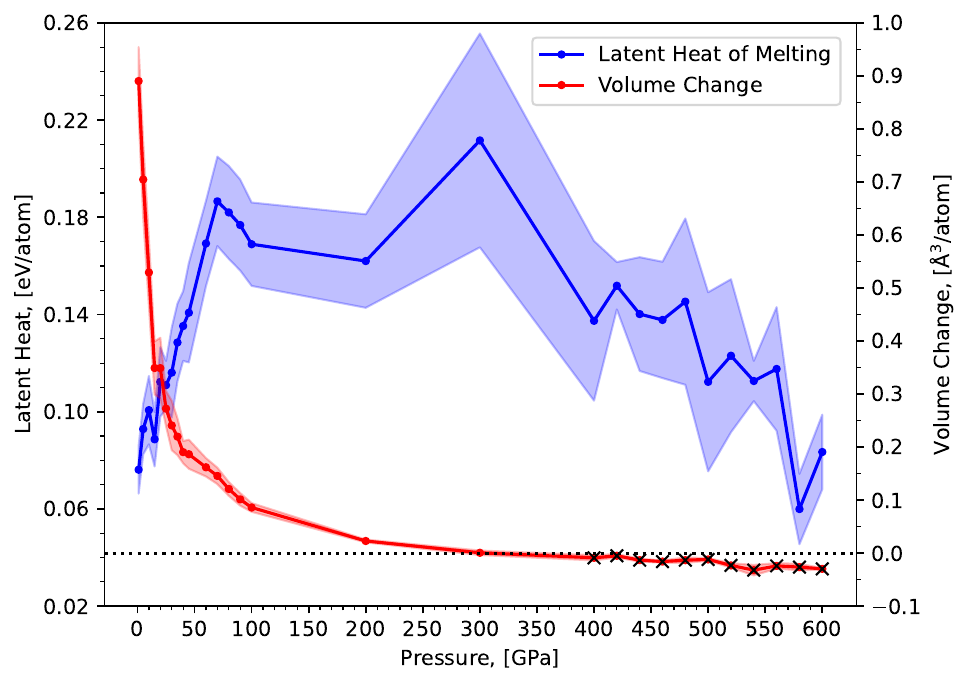}
\end{center}
\vspace{-20pt}
\caption{\textbf{Volume changes and latent heats of melting for magnesium from 1-600~GPa.} Results are collected from the enthalpy curves produced by sampling the C4~O4~D14 ACE model and are shown in Figure~\ref{fig:enth_curves}. Pale regions indicate the STD associated with the measurement, as explained in Appendix~\ref{sec:lh_and_enths}, and the black crosses indicate negative thermal expansion.}
\label{fig:latent_heats}
\end{figure}

With access to the temperature dependent enthalpy curves across the entire pressure range from \gls{ns}, we can easily compute the latent heats of melting as a function of pressure. Since the temperature dependent enthalpy curves are noisy, we approximate these quantities by fitting a linear function before and after the transition temperature and then calculating the difference between these functions at the transition temperature. This removed some of the noise, however, the temperature range the functions are fitted to affects the latent heat value and so we perform multiple fits, with a buffer range of 100-2000~K, and use the mean and \gls{std} of values to provide error bars to the measurements shown in Figure~\ref{fig:latent_heats}.

\begin{figure*}
\begin{center}
\includegraphics[width=\linewidth,angle=0,scale=0.95]{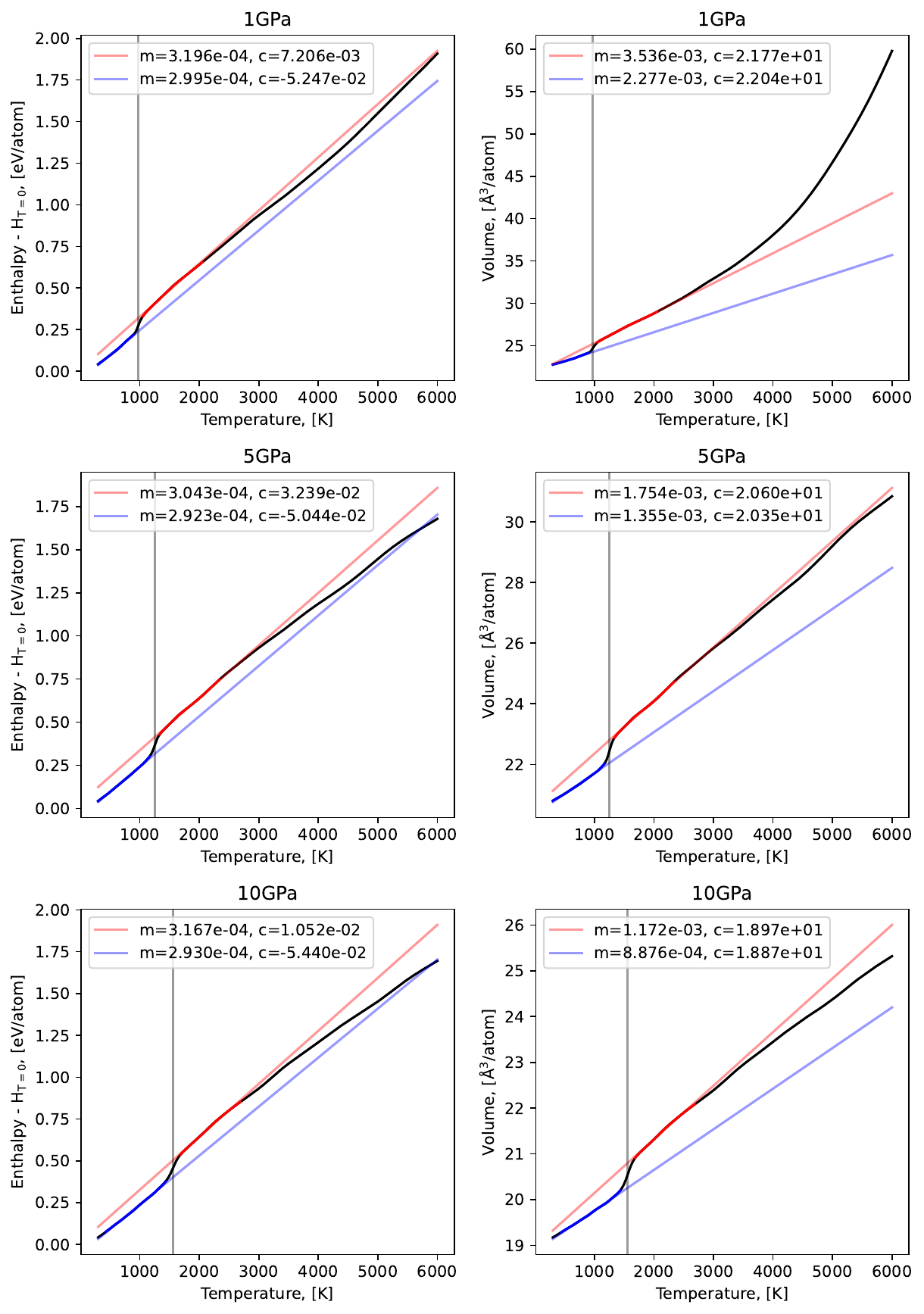}
\end{center}
\end{figure*}
\begin{figure*}
\begin{center}
\includegraphics[width=\linewidth,angle=0,scale=0.95]{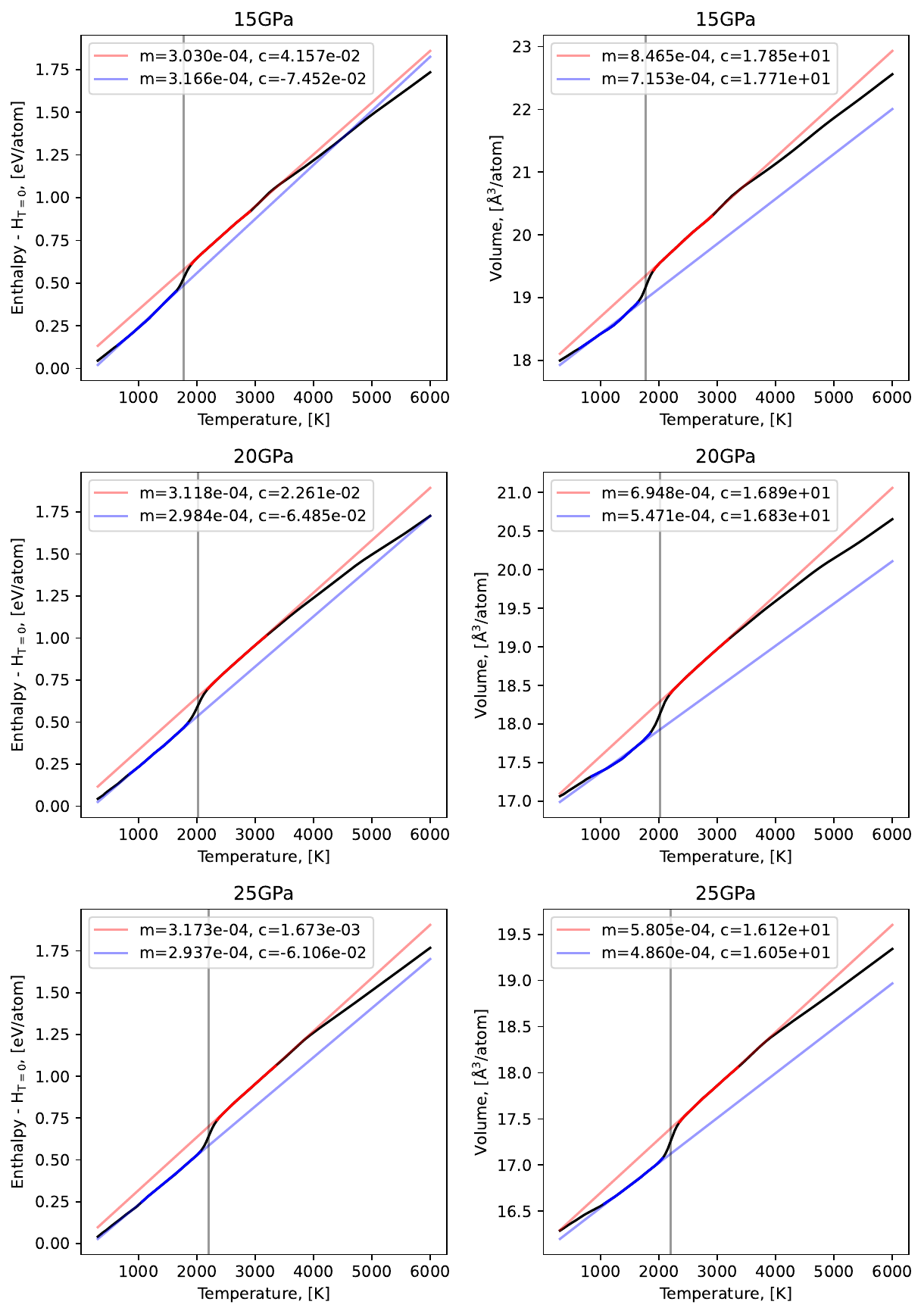}
\end{center}
\end{figure*}
\begin{figure*}
\begin{center}
\includegraphics[width=\linewidth,angle=0,scale=0.95]{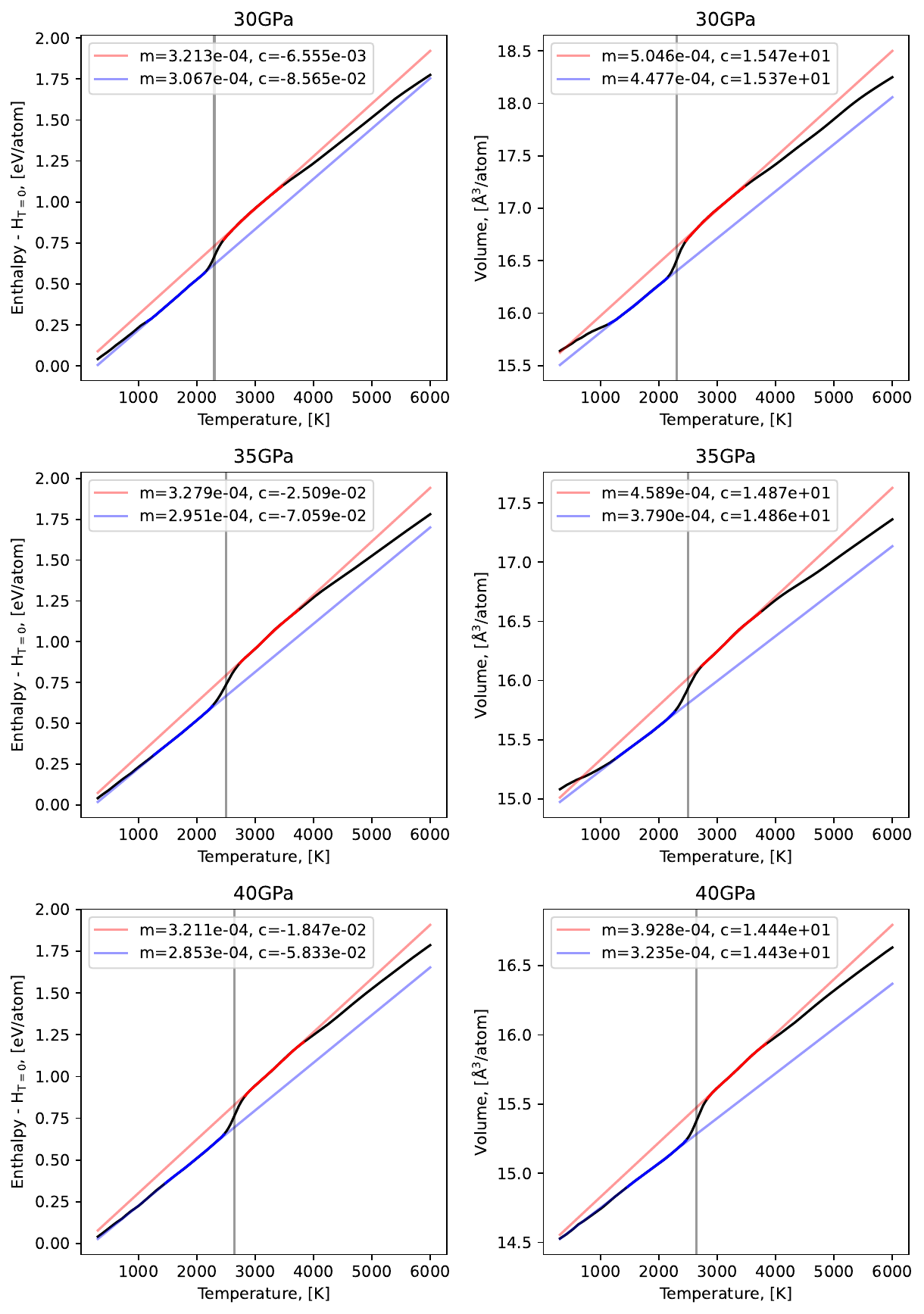}
\end{center}
\end{figure*}
\begin{figure*}
\begin{center}
\includegraphics[width=\linewidth,angle=0,scale=0.95]{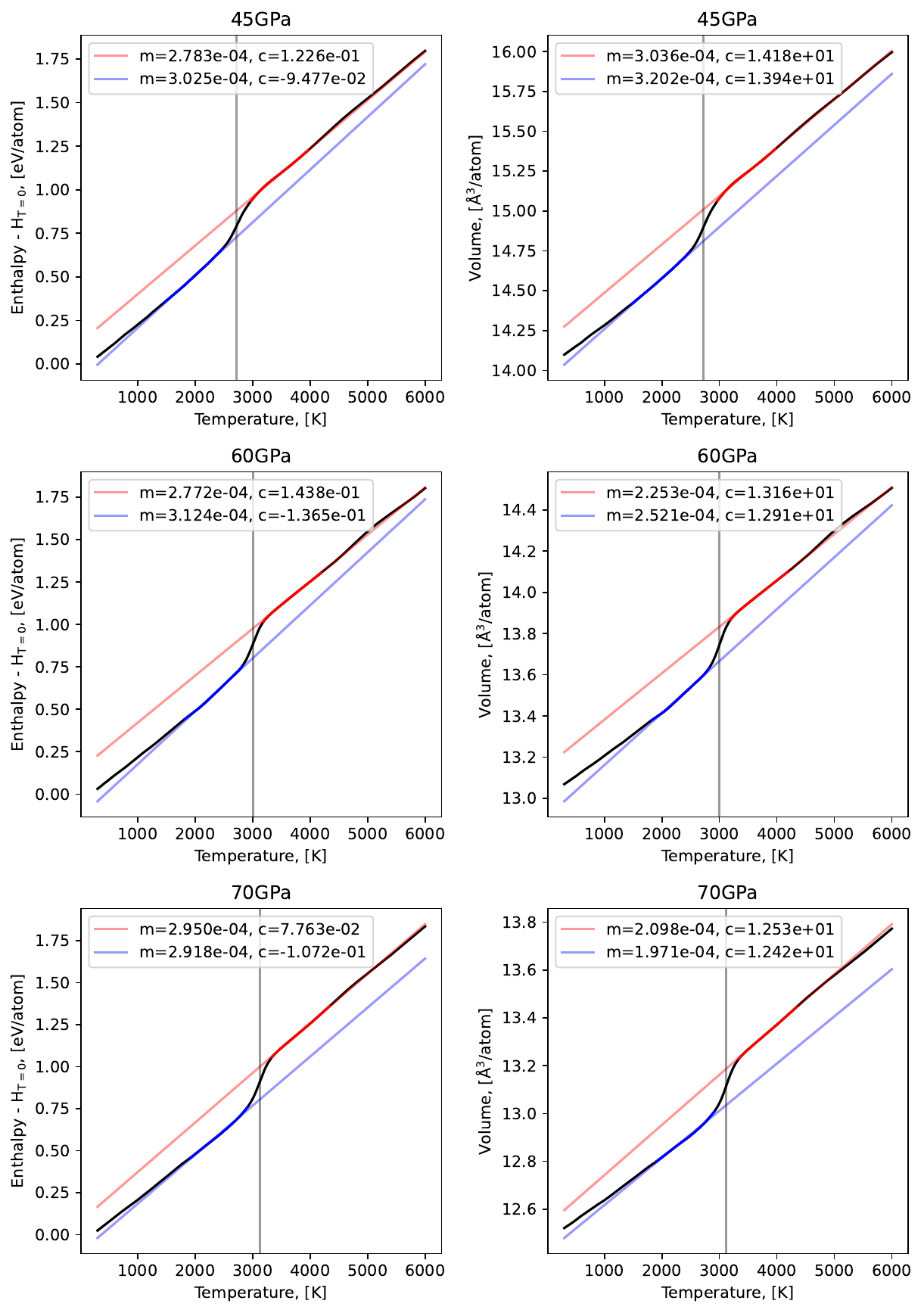}
\end{center}
\end{figure*}
\begin{figure*}
\begin{center}
\includegraphics[width=\linewidth,angle=0,scale=0.95]{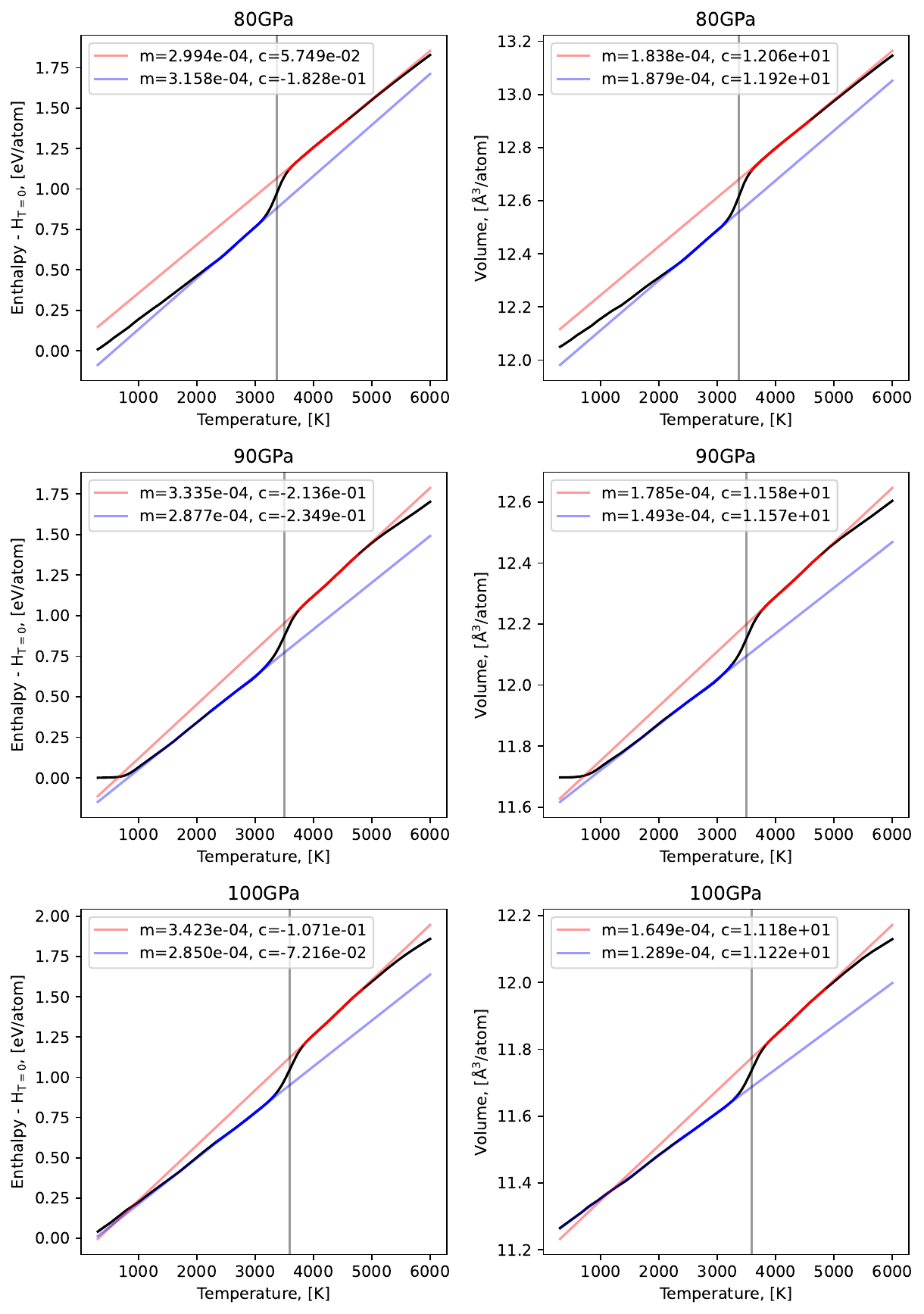}
\end{center}
\end{figure*}
\begin{figure*}
\begin{center}
\includegraphics[width=\linewidth,angle=0,scale=0.95]{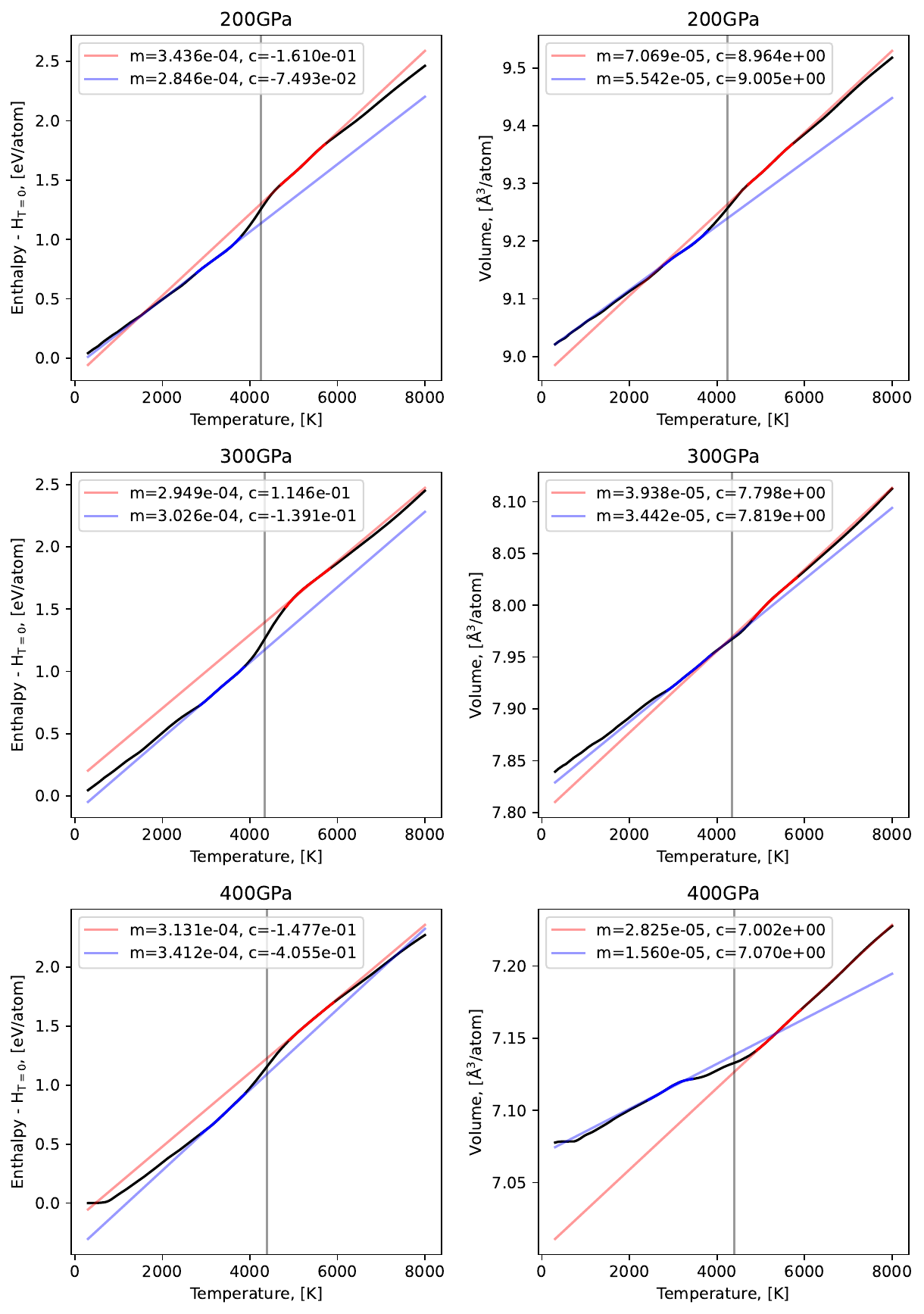}
\end{center}
\end{figure*}
\begin{figure*}
\begin{center}
\includegraphics[width=\linewidth,angle=0,scale=0.95]{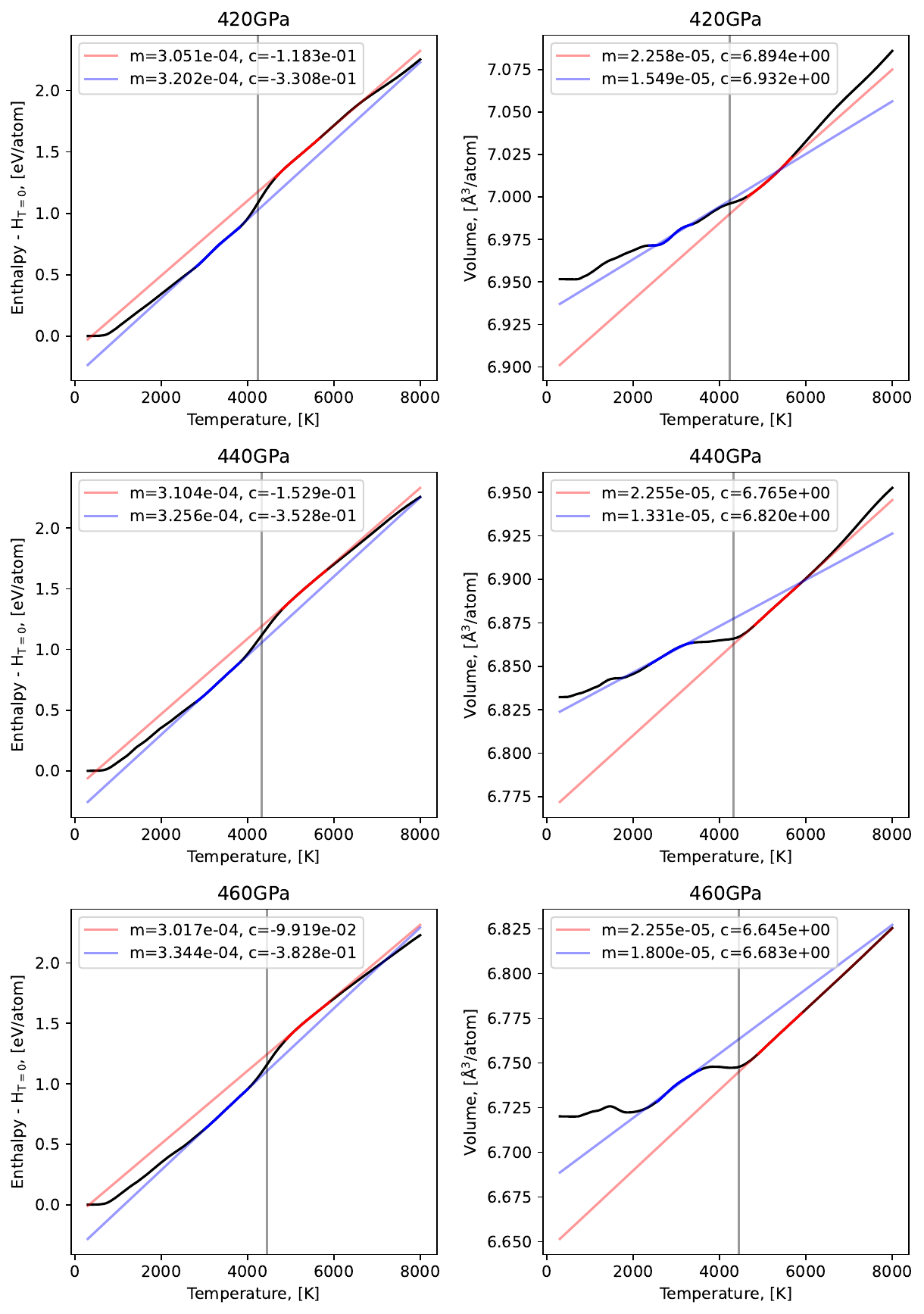}
\end{center}
\end{figure*}
\begin{figure*}
\begin{center}
\includegraphics[width=\linewidth,angle=0,scale=0.95]{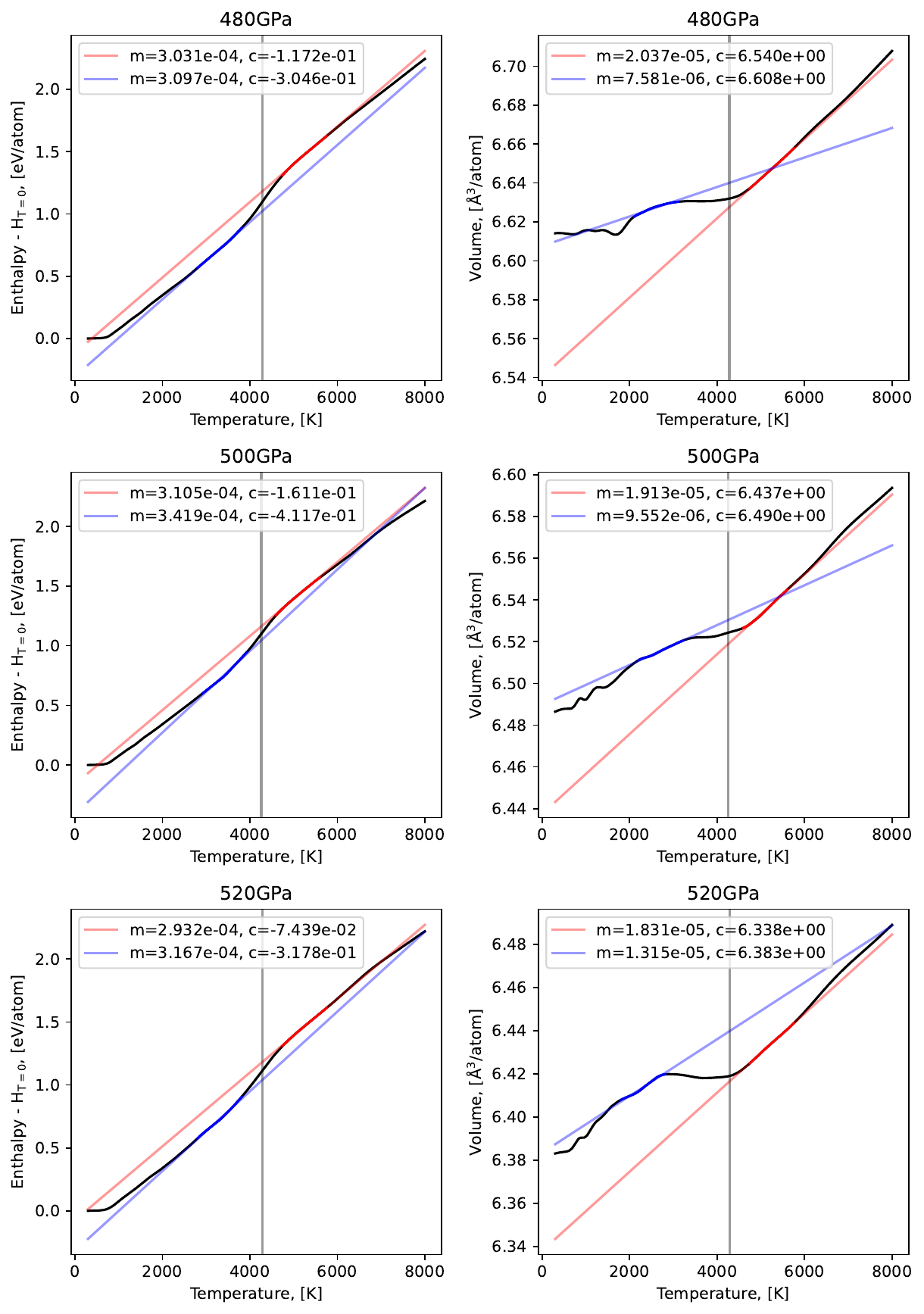}
\end{center}
\end{figure*}
\begin{figure*}
\begin{center}
\includegraphics[width=\linewidth,angle=0,scale=0.95]{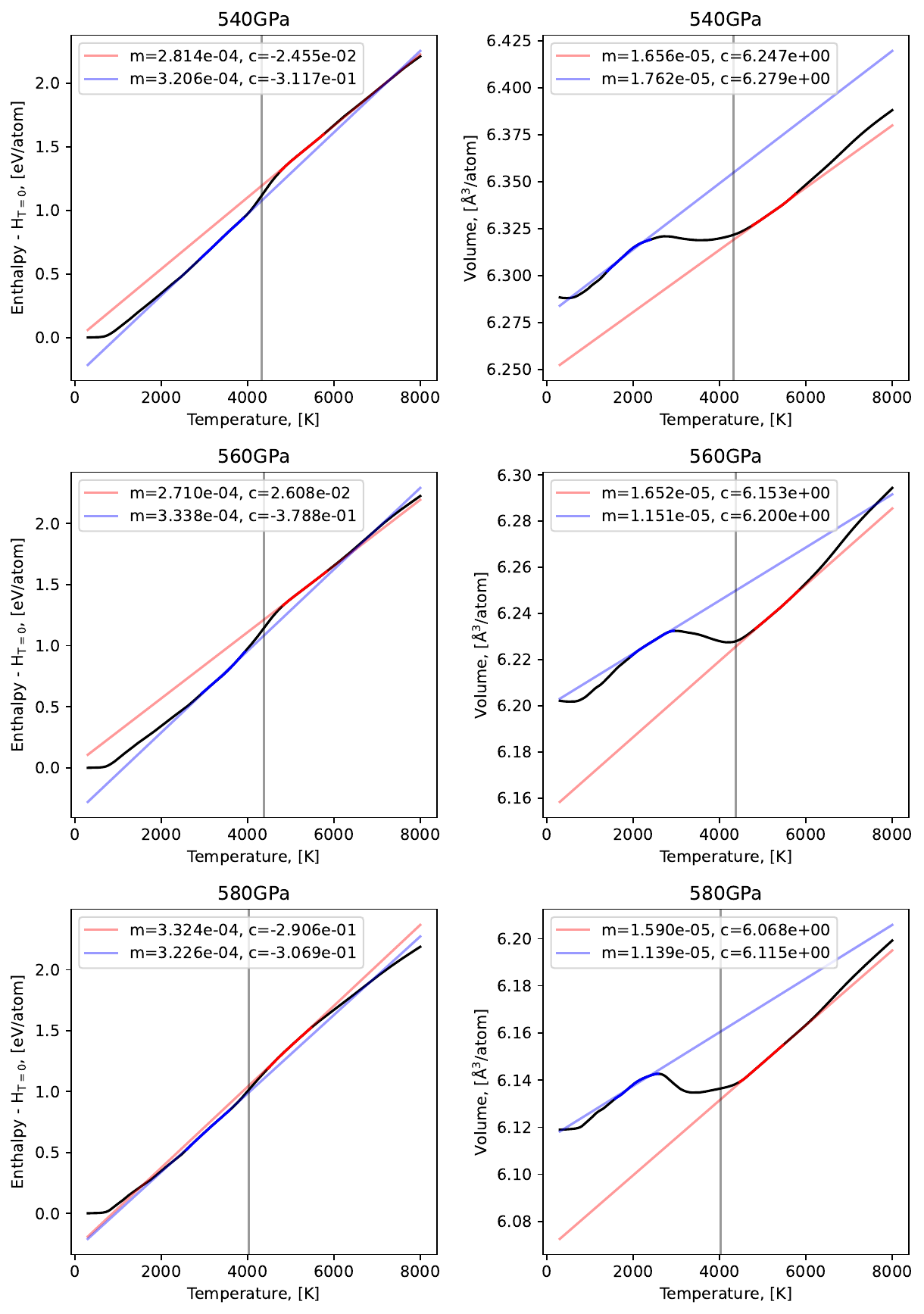}
\end{center}
\end{figure*}
\begin{figure*}
\begin{center}
\includegraphics[width=\linewidth,angle=0,scale=0.95]{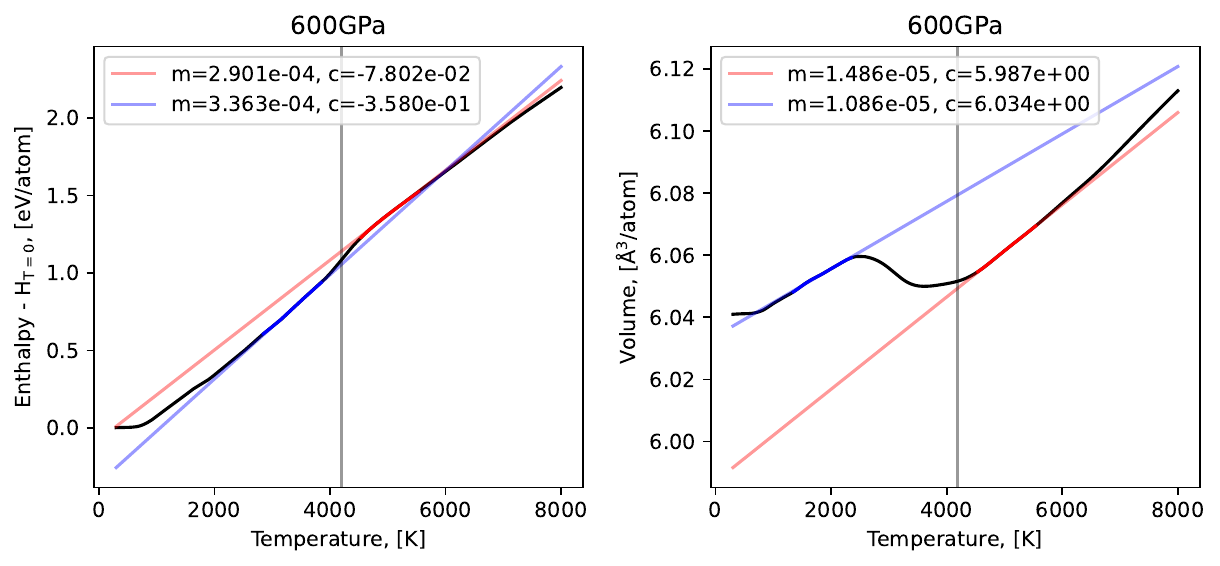}
\end{center}
\vspace{-20pt}
\caption {\textbf{Temperature-enthalpy and temperature-volume plots produced from 64-atom NS using the C4~O4~D14 model at 1-600~GPa.} The transition temperatures are shown as vertical black lines. Linear functions were fitted to their respective coloured regions of the plots with the gradient, m, and y-intercept, c, given in the legend. This data was used to produce the results shown in Figure~\ref{fig:latent_heats} and calculate the thermal expansion coefficients shown in Figure~\ref{fig:thermal_exp}.}
\label{fig:enth_curves}
\end{figure*}

\end{document}